\RequirePackage{fix-cm}
\documentclass[
  aps,
  prl,
  10pt,
  reprint,
  superscriptaddress,
  amsmath,
  amssymb,
  longbibliography
]{revtex4-2}

\ifdefined\pdfminorversion
\fi

\usepackage[utf8]{inputenc}
\usepackage{graphicx}
\usepackage{dcolumn}
\usepackage{bm}
\usepackage{etoolbox}
\usepackage{hyperref}
\usepackage{CJKutf8}
\usepackage{placeins}

\begin{document}

\title{Quantum Metric Control of Topological Superconductivity and Incommensurate Charge Density Wave in an Interacting Lieb-Kitaev Chain Model}

\author{Jiale Xuan}
\affiliation{Key Laboratory of Artificial Structures and Quantum Control (Ministry of Education), School of Physics and Astronomy, Shanghai Jiao Tong University, Shanghai 200240, China}
\author{Grant Z. X. Yang}
\affiliation{Department of Physics, Hong Kong University of Science and Technology, Clear Water Bay, Hong Kong, China}
\author{K. T. Law}
\email{phlaw@ust.hk}
\affiliation{Department of Physics, Hong Kong University of Science and Technology, Clear Water Bay, Hong Kong, China}
\author{Xiao Yan Xu}
\email{xiaoyanxu@sjtu.edu.cn}
\affiliation{Key Laboratory of Artificial Structures and Quantum Control (Ministry of Education), School of Physics and Astronomy, Shanghai Jiao Tong University, Shanghai 200240, China}

\date{\today}

\begin{abstract}
We study an interacting spinless Lieb-Kitaev chain model in the nearly flat-band regime, motivated by the recent observation that Majorana zero modes in the noninteracting Lieb-Kitaev system inherit two decay modes: an ordinary Bogoliubov-de Gennes (BdG) length and a longer geometric length controlled by the quantum metric. By tuning both the interaction strength and quantum metric, we identify a topological superconducting (TSC) phase, an incommensurate charge-density-wave (ICDW)phase, and a commensurate charge-density-wave (CDW) phase. Inside the TSC phase, the QML is enhanced when increasing the interactions. In the ICDW phase, the interactions are modulated by quantum metric to generate an effective chemical potential, leading to a discontinuous incommensurate-to-commensurate (IC-C) transition. We verify these phases and conclusions using low energy effective theory and density-matrix renormalization group (DMRG) calculations.
\end{abstract}

\maketitle
\setlength{\abovedisplayskip}{4pt plus 1pt minus 1pt}
\setlength{\belowdisplayskip}{4pt plus 1pt minus 1pt}
\setlength{\abovedisplayshortskip}{2pt plus 1pt}
\setlength{\belowdisplayshortskip}{3pt plus 1pt minus 1pt}
\setlength{\jot}{1pt}
\makeatletter
\let\PRLoriginalcaption\@makecaption
\patchcmd{\@makecaption}{\small\rmfamily}{\fontsize{8.5}{10}\selectfont\rmfamily}{}{}
\makeatother

\noindent\textit{Introduction.}  Over the past two decades, flat-band materials have attracted considerable interest. 
Owing to the quenching of kinetic energy in flat bands, interaction effects are substantially enhanced and 
can dominate the low-energy physics, rendering flat-band systems a fertile platform for investigating strongly 
correlated phenomena and interaction-driven quantum phases. This has led to the realization of a wide range of 
exotic states, including unconventional superconductivity~\cite{caoUnconventionalSuperconductivityMagicangle2018,peottaSuperfluidityTopologicallyNontrivial2015a,yankowitzTuningSuperconductivityTwisted2019},
magnetism ~\cite{repellinFerromagnetismNarrowBands2020,linFlatbandsEmergentFerromagnetic2018}, 
and fractional topological phases ~\cite{tangHighTemperatureFractionalQuantum2011,regnaultFractionalChernInsulator2011,royBandGeometryFractional2014}. 

For a superconducting (SC) phase within a conventional single-band BCS picture, a perfectly flat band has vanishing band 
curvature and thus an infinite effective mass would appear to strongly inhibit superconductivity. 
In contrast, in multiband flat-band systems, the complete quenching of the band dispersion does not eliminate 
the nontrivial momentum-space structure of the Bloch wave functions
~\cite{huhtinenRevisitingFlatBand2022,tormaQuantumMetricEffective2018a,huGeometricConventionalContributions2025,tovmasyanEffectiveTheoryEmergent2016a}.
 The Bloch states can retain a finite extent in Hilbert space across the Brillouin zone, 
 reflecting their variation with crystal momentum, and this quadratic spread is quantified by the quantum metric 
 through the infinitesimal distance between neighboring Bloch states~\cite{restaInsulatingStateMatter2011,chengQuantumGeometricTensor2010,yuQuantumGeometryQuantum2025}. 
 Such nontrivial quantum geometry provides an additional channel for superconducting transport that is absent in 
 the conventional single-band picture. As a consequence, a finite superfluid weight and robust superconducting 
 coherence can persist even in the flat-band limit, despite the absence of conventional band-dispersion contributions
 ~\cite{peottaSuperfluidityTopologicallyNontrivial2015a,xieTopologyBoundedSuperfluidWeight2020a,tormaSuperconductivitySuperfluidityQuantum2022a,julkuGeometricOriginSuperfluidity2016a}.

The superconducting coherence length is a fundamental length scale that characterizes the spatial extent of 
superconducting correlations. In conventional BCS superconductors, 
it is given by \(\xi_{\mathrm{BCS}}\sim \hbar v_F/\Delta\), where \(v_F\) is the Fermi velocity and \(\Delta\) 
is the pairing gap. This length scale measures the characteristic size of Cooper pairs. By comparison, recent
developments have shown that the quantum metric of Bloch wave functions not only generates a finite geometric 
contribution to the superfluid weight, but also establishes an intrinsic length scale governing superconducting
phenomena~\cite{chenGinzburgLandauTheoryFlat2024,huAnomalousCoherenceLength2025,yang2026quantumgeometrickondocloud}.
This geometric viewpoint became significant in the recent analysis of Majorana zero modes in the noninteracting 
Lieb--Kitaev model ~\cite{guoMajoranaZeroModes2025}. There, the edge mode exhibits two distinct correlation lengths: a short Bogoliubov--de Gennes scale associated with
ordinary topological superconducting localization ~\cite{kitaevUnpairedMajoranaFermions2001,miaoMajoranaZeroModes2018}, 
and a longer geometric scale inherited from the spread of flat-band wave functions. The latter is controlled by the 
quantum metric and can become parametrically large in the nearly flat-band regime. This immediately raises a  many-body
question: once repulsive interactions are introduced, does the geometric long tail survive as a physically meaningful
scale, or is it washed out by correlation-driven reconstruction?

In previous studies, interactions in flat band models were typically introduced in the form of an attractive pairing channel, 
which directly promotes Cooper pairing and thereby induces superconductivity.
Representative examples in one dimension include stub~\cite{thuminFlatBandSuperconductivity2023}, 
sawtooth~\cite{chanPairingSuperconductivityQuasi2022} and Creutz~\cite{chanPairingSuperconductivityQuasi2022,mondainiPairingSuperconductivityFlat2018,tovmasyanEffectiveTheoryEmergent2016a,salernoDrudeWeightManybody2023} model.
Here, we retain the attractive interaction responsible for superconducting
pairing while introducing a repulsive interaction to incorporate competing correlation effects beyond the 
pairing channel. Such a coexistence of attractive and repulsive interactions occurs in many
electronic systems: an effective attractive pairing interaction may arise from phonons, collective fluctuations,
or superexchange processes, while Coulomb interactions remain repulsive. This situation is encountered,
for example, in electron-phonon superconductors~\cite{morelCalculationSuperconductingState1962,sacepeQuantumBreakdownSuperconductivity2020}, 
moiré flat-band systems~\cite{wuTheoryPhononMediatedSuperconductivity2018,lianTwistedBilayerGraphene2019}.
and extended Hubbard models~\cite{pencPhaseDiagramOnedimensional1994,aligiaPhaseDiagramOnedimensional2000}.

To address this question, we employ DMRG
calculations and develop a low-energy effective theory 
inspired by Refs.~\cite{giamarchi2004,gangadharaiahMajoranaEdgeStates2011a} 
to investigate the interaction effects. The results show 
transitions among a topological superconducting (TSC) phase,
an incommensurate charge-density-wave (ICDW) phase, and a charge-density-wave (CDW) phase.
An incommensurate phase associated with a doped Luttinger liquid has been observed in studies of the Kitaev--Hubbard
chain~\cite{mahyaehStudyPhaseDiagram2020,rahmaniPhaseDiagramInteracting2015,stoudenmireInteractionEffectsTopological2011a}. In the TSC phase, we apply a mean-field
analysis and show that flat-band geometry contributes to the
correlation length and can influence Majorana instability. Our continuum model shows that 
the ICDW phase originates from
the interaction and geometry induced effective chemical potential generated
by the multiorbital projection. It is the competition between this effective
chemical potential and the half-filling Umklapp process that drives the IC-C 
transition. In particular, the process differs from the traditional IC-C transition
where doping at the fixed point approaches zero.

\par\smallskip\noindent\phantomsection\label{sec:model}\textit{Model and theoretical framework.}---
The hamiltonian of the interacting Lieb--Kitaev chain model is:
\begin{gather}
\begin{aligned}
H_0={}&\sum_{j,\alpha=A,B,C}
\bigl[-t c_{j\alpha}^{\dagger}c_{j+1,\alpha}
+\Delta c_{j\alpha}c_{j+1,\alpha}+\mathrm{h.c.}\bigr]
\\
&+V\sum_j
\bigl(c_{jA}^{\dagger}c_{jA}-c_{jC}^{\dagger}c_{jC}\bigr)
\\
&-J\sum_j\bigl[
(c_{jA}^{\dagger}+c_{jC}^{\dagger})
(c_{jB}+c_{j+1,B})+\mathrm{h.c.}\bigr].
\end{aligned}
\label{h0}\\
H_U=U\sum_{j=1}^{L-1}\sum_{\alpha=A,B,C}
\bigl(\hat n_{j\alpha}-\tfrac12\bigr)
\bigl(\hat n_{j+1,\alpha}-\tfrac12\bigr).
\label{Uinteraction}
\end{gather}

\begin{figure}[t]
    \centering
    \includegraphics[width=\columnwidth]{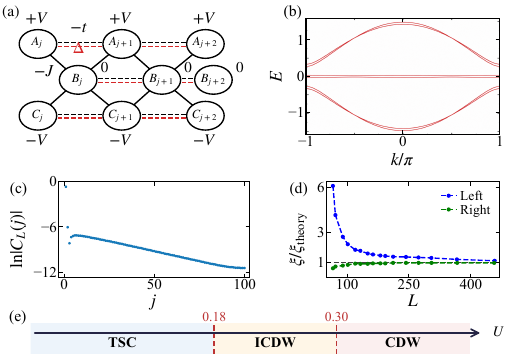}
    \caption{
    (a) Lattice structure of the Lieb--Kitaev chain.
    (b) Energy bands for $V=0.6$, 
    and $\Delta=0.018$.
    (c) Two decay lengths of the left Majorana
    $C_L(j)=\langle c_{jB}\gamma_L\rangle$, where
    $\gamma_L=c_{1B}+c_{1B}^{\dagger}$.
    (d) Finite-size effects of decay lengths $\xi$ extracted from the left-boundary
    (blue) and right-boundary (green) correlation profiles
    normalized by $\xi_{\mathrm{theory}}=4\xi_{\mathrm{QM}}$.
    Here the parameters are $a=3$, and $V=0.1$, the geometric
    reference scale is
    $4\xi_{\mathrm{QM}}\simeq\sqrt{2}Ja/V\approx42.4$. 
        (e) Phase diagram as a function of $U$, showing
    the TSC, ICDW, and CDW regimes, with transition points
    indicated at $U=0.18$ and $U=0.30$.
    }
    \label{fig:1}
\end{figure}

The model describes a one-dimensional three-orbital lattice under open boundary conditions
(Fig.~\ref{fig:1}(a)). We denote the
number of unit cells by $L$ and the total number of orbitals by $N=3L$. The onsite energies of
the $A$, $B$, and $C$ orbitals are shifted by \(+V\), \(0\), and \(-V\), respectively. The hopping amplitude \(J\) couples
the \(B\) orbital to the neighboring \(A\) and \(C\) orbitals, both within the same unit cell
and between adjacent unit cells, forming the characteristic Lieb-like lattice geometry. To
realize a TSC phase, nearest-neighbor intra-orbital pairing with
amplitude \(\Delta\) is introduced between the same orbitals in adjacent unit cells. 
To maintain the flat-band regime, the dispersion of the middle band, characterized by the energy 
scale \(2t\), should be smaller than the separation from the dispersive bands, \(E_g=|V|\) 
and the dominant hopping $J$. For the parameters $t=0.03  , J=1 , \Delta=0.018$, with 
$V$ varied over $0.1-0.6$,
the flat-band condition remains satisfied (Fig.~\ref{fig:1}(b)). Nevertheless, 
the dispersion is small but non-zero, allowing the renormalization of the Tomonaga--Luttinger-type
continuum model. 

\begin{figure*}[t]
    \centering
    \includegraphics[width=\textwidth]{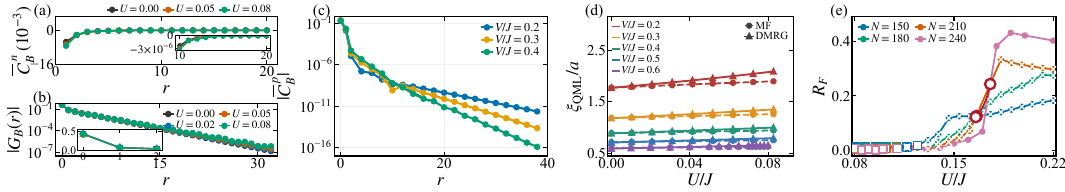}
    \caption{TSC phase.
    (a) Connected $BB$ density correlation at $V=0.6$.
    (b) Single-particle correlation  $G_B(r)=\langle c_{iB}^{\dagger}c_{i+r,B}\rangle$
    at $V=0.6$.
    (c) Connected pair correlation for different QMLs at
     $U\simeq0.02$.
    (d) Interacting QML versus $U$, in units of the unit-cell spacing $a$.
    Circles with dashed lines show $\xi_{\mathrm{QML}}^{\mathrm{MF}}$;
    triangles with solid lines show $\xi_{\mathrm{QML}}^{\mathrm{DMRG}}=\xi_G^{\mathrm{fit}}/4$,
    extracted from $G_B(r)$. Colors distinguish $V=0.2,0.3,0.4,0.5,0.6$.
    (e) Fourier correlation ratio for the TSC--ICDW transition, $R_F=1-|F(Q-2\pi/L)|/|F(Q)|$ with $F(r)=r^2\overline C_B(r)$, at $V=0.6$.
    The red curve connects finite-size crossings,
    extrapolating to $U_c\simeq0.18$
    }
    \label{fig:2}
\end{figure*}

Recent work~\cite{guoMajoranaZeroModes2025} showed that the Majorana edge profile in the noninteracting
Lieb--Kitaev model inherits both a short localization length and a longer scale tied to $\xi_{\rm QM}$.
On the one hand, solving the model directly can give the expression of $\xi_{\rm QM}$ [Eq.~\eqref{eq:IQML-bare}]. On the other hand, the Majorana wave function under
open boundary conditions can be constructed from the
projected bulk Green’s function, $\psi_L(x)\propto G^p(x,x_{\rm L};0)$, where
\begin{equation}
    G^p(x,x';0)
    =
    \sum_{n=\pm}
    \int\frac{dk}{2\pi}\,
    \frac{g_n(k)g_n^\dagger(k)}
    {-\epsilon_n(k)}
    e^{ik(x-x')}.
\end{equation}
$\epsilon_n(k)$ denotes the dispersion of the two low-energy
Bogoliubov quasiparticle bands, while $g_n(k)$ is the corresponding
BdG Bloch eigenstate. The quantum metric length originates from the momentum-dependent
orbital composition of the Bloch states $g_n(k)$, which gives the projected
flat-band orbitals a finite spatial extent of $4\xi_{\rm QM}$ even when the band
dispersion vanishes. We examined the two modes through the exponential decay lengths of single-particle correlation functions and correlation functions. The numerical results 
are consistent with theoretical results in the thermodynamic limit (Fig.~\ref{fig:1}(c,d)).

We introduce a density interaction between neighboring unit cells into this model, as given in Eq.~\eqref{Uinteraction}
and set the bare chemical potential to zero, so that no external chemical-potential bias is
imposed. Our results exhibit a TSC phase, an ICDW phase and a CDW phase (Fig.~\ref{fig:1}(e)).
The numerical methods are described in Sec.~\ref{numeriacal details} of the Supplemental Material.


\par\smallskip\noindent\textit{Regime I: topological regime with quantum metric}---
The weak-interaction regime labelled TSC 
is characterized by spatially uniform bulk observables and a
noncritical entanglement structure.
The normal bond expectation values (Fig.~\ref{fig:S3})
$\langle c^\dagger_{i\alpha}c_{i+1,\alpha}\rangle$ and pairing
expectation values $\langle c_{i\alpha}c_{i+1,\alpha}\rangle$
likewise approach uniform bulk values for $\alpha=A,B,C$.
Thus, no bulk CDW, intraorbital bond-order-wave (BOW), or pair-density-wave (PDW) modulation is resolved
in these profiles. The entanglement entropy reaches an essentially size-independent
plateau for different sizes.
Fitting the central half of the chain yields $c_{\mathrm{eff}}\simeq0$, consistent with a noncritical bulk.

The single-particle correlation function on the $B$ orbital, $G_B(r)=\langle c^\dagger_{iB}c_{i+r,B}\rangle$
exhibits approximately exponential decay
(Fig.~\ref{fig:2}(b)) with $\xi\simeq2.38, 2.44, 2.53, 2.63$ unit cells for $a=3$ at
$U=0,0.02$, $0.05,0.08$, respectively. This is similar to density (Fig.~\ref{fig:2}(a)), bond, and pair (Fig.~\ref{fig:2}(c)) correlation functions, where
$P_B(i)=c_{iB}c_{i+1,B}$ and
$C_B^p(i,j)=\langle P_B^\dagger(i)P_B(j)\rangle
-\langle P_B^\dagger(i)\rangle\langle P_B(j)\rangle$.
We observe an increase in the interacting QML with increasing $U$, 
and a decrease in the interacting QML with increasing $V$, 
which represents $\xi_{\rm QM}$. These trends motivate a Hartree mean-field description
of the interacting QML (Sec.~\ref{app:interacting-QML} of the Supplemental Material).
The orbital-dependent Hartree shifts renormalize the
splitting as
$V_{\mathrm{eff}}=V-U(\bar n_C-\bar n_A)$,
where $\bar n_\alpha$ denotes the bulk occupation of
orbital $\alpha$ obtained from DMRG. To leading order,
we estimate
\begin{equation}
\xi_{\mathrm{QML}}^{\mathrm{MF}}
\simeq
\xi_{\mathrm{QML}}^{(0)}(V)
-U(\bar n_C-\bar n_A)
\frac{\partial\xi_{\mathrm{QML}}^{(0)}}{\partial V}.
\end{equation}
Since $\partial_V\xi_{\mathrm{QML}}^{(0)}<0$,
the orbital polarization $\bar n_C>\bar n_A$ increases
the predicted geometric length. We compare this estimate
with the length inferred from the single-particle
correlations in Fig.~\ref{fig:2}(d). Our approximation 
agrees with the DMRG results to within \(10\%\). The
Majorana wave function can be expressed as a sum of two decay modes:
\begin{equation}
\Psi_L(x)
\simeq
\mathbf A_{L,g}
e^{-x/\xi_{\mathrm g}}
+\mathbf A_{L,s}
e^{-x/\xi_{\mathrm s}}
\end{equation}
When the geometric component dominates the Majorana
tail, the same length controls the scale of Majorana overlap,
$|t_{RL}|_{\mathrm{env}}\propto
e^{-L/(4\xi_{\mathrm{QML}}^{\mathrm{MF}})}$,
with $L$ the physical chain length.
The interaction-induced growth of the geometric length
therefore provides a mechanism for stronger hybridization, thus weakening the
finite-size protection of Majorana degeneracy and resulting in
a lower critical interaction strength for the TSC--ICDW transition.
\begin{figure*}[t]
    \centering
    \includegraphics[width=\textwidth]{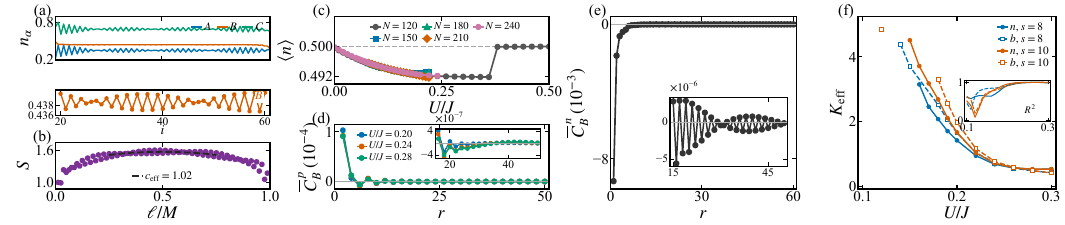}
    \caption{
    ICDW phase. $U/J=0.24$ for (a,b,e), $N=240$ for (a,b,d,e,f), $V=0.6$
    (a) Local occupations $\langle n_{i\alpha}\rangle$ showing oscillations.
    (b) Intercell von Neumann entropy for the same state. The dashed line
    fits $S(\ell)=(c_{\rm eff}/6)\ln[(2M/\pi)\sin(\pi\ell/M)]+s_0$
    over $20\leq\ell\leq60$, yielding $c_{\rm eff}=1.02$;
    $M=N/3=80$. (c) Average occupation $\langle N_{\rm tot}\rangle/N$, showing
    finite doping that is not attributable to finite-size effects in the ICDW regime and
    half filling in the CDW regime. (d) Connected pair correlation,
    without long-range or quasi-long-range order.
    (e) Connected density correlation, with an oscillatory tail.
    (f) $B$-orbital effective Luttinger
    parameters from density and bond profile fits, using fitting windows $s<r<L-s$ with $s=8,10$.
    }
    \label{fig:3}
\end{figure*}

The density correlations decay exponentially in the TSC
phase (Fig.~\ref{fig:2}(a)), whereas the ICDW phase exhibits power
decaying correlations with an oscillatory tail.
We extract the characteristic wave vector $Q$ from the
Fourier spectrum of this tail and use it to construct
the correlation ratio.
We locate the TSC--ICDW transition by extrapolating
the crossings of the correlation-ratio curves for
different system sizes to the thermodynamic limit,
obtaining $U_c\simeq0.18$ for $V=0.6$ as shown in
Fig.~\ref{fig:2}(e) and $U_c\simeq0.15$ for $V=0.3$ (Fig.~\ref{fig:S2}).
The trend in the transition point is consistent with
our mean-field picture: smaller $V$ corresponds to a
larger interacting QML, which enhances Majorana
hybridization and weakens finite-size protection.

\par\smallskip\noindent\textit{Regime II: ICDW regime.}---
The second regime is labelled ICDW. Fitting the entanglement entropy to the Calabrese--Cardy (CC) formula for open boundary conditions (OBCs)~\cite{pasqualecalabreseEntanglementEntropyQuantum2004} yields
a central charge $c\simeq1$, consistent with a gapless state (Fig.~\ref{fig:3}(b)).
Figure~\ref{fig:3}(a) displays spatial oscillations in contrast to the uniform bulk profiles
of the TSC regime. Corresponding oscillations occur in both
the bond and pairing channels. The connected correlations, however, exhibit exponential decay in the pairing 
channel (Fig.~\ref{fig:3}(d)) and power-law decay with an oscillatory tail in the density channel (Fig.~\ref{fig:3}(e)). This is 
known in a Gaussian Tomonaga--Luttinger theory as Friedel 
oscillations~\cite{Giamarchi1997,Voit1995,eggerFriedelOscillationsLuttinger1996}
~\eqref{eq:app_density_corr_form}
\begin{equation}
C_B^{\mathrm{bulk}}(r)
\simeq
-\frac{D_B}{r^2}
+\frac{A_C\cos(Qr)}{r^{2K}},
\qquad
D_B=\frac{Z_B^2K}{2\pi^2},
\end{equation}
where $K$ is the Luttinger parameter, $Z_B$ is the projection onto the B orbital
and $Q$ is the doping. $Q$ is derived from DMRG data of $\langle N_{\rm tot}\rangle$
(Fig.~\ref{fig:3}(c)), and the doping is stable against finite-size effects.
Therefore, we identify a gapless ICDW phase and the $IC-C$ transition point
is near $U=0.3$ for $V=0.6$. The oscillations in the pair channel
 are induced by charge-density modulations and do not indicate 
 genuine long-range or quasi-long-range superconducting order.

To interpret the numerical results, we construct a low-energy
effective theory (Sec.~\ref{app:lowenergy} of the Supplemental Material) by projecting the  Lieb--Kitaev
Hamiltonian in Eqs.~\eqref{h0} and~\eqref{Uinteraction} onto its narrow middle band.
Linearizing the projected dispersion near the two  Fermi
points and following standard bosonization yields the effective Hamiltonian [Eq.~\eqref{eq:app-effective-Hamiltonian-expanded}]:

\begin{gather}
\begin{aligned}
H_{\mathrm{eff}}=\int_0^\ell dx\,\Bigl\{
&\frac{v}{2\pi}\bigl[
K(\partial_x\theta)^2+K^{-1}(\partial_x\phi)^2\bigr]
\\
&+\frac{\mu_{\mathrm{eff}}}{\pi}\partial_x\phi
-\frac{g_u}{2\pi^2\alpha_0^2}:\cos[4\phi(x)]:_0
\\
&-\frac{2\Delta}{\pi\alpha_0}:\cos[2\theta(x)]:_0
\Bigr\}.
\end{aligned}
\label{lowenergy}\\
K=\sqrt{\frac{2\pi v_F+g_4-g_2}{2\pi v_F+g_4+g_2}},
\;
v=\frac{\sqrt{(2\pi v_F+g_4)^2-g_2^2}}{2\pi}.
\end{gather}

Here $\alpha_0$ is the short-distance cutoff, chosen as the unit-cell spacing, $\alpha_0=a$. The noninteracting part yields a
Tomonaga--Luttinger Hamiltonian characterized by two dual fields $(\phi,\theta)$ with
parameters $K$ and $v$. The scattering term in the projected interaction generates
interband and intraband scattering $g_2$ and the Umklapp process $g_u$ near
half filling. The bilinear term in the projected interaction contributes
to a dispersion perturbation $g_4$ and a chemical-potential shift $\mu_{\mathrm{eff}}$, although the bare chemical potential
remains zero. In particular,
the ICDW phase originates from this shift [Eq.~\eqref{eq:app-bilinear-coefficients}]:
\begin{equation}
\mu_{\mathrm{eff}}=U(1-\Sigma_{\mathrm{mid},F}-\Sigma_{\mathrm{bg},F}),
\end{equation}
where $\Sigma_F$ is the bilinear contribution to the self-energy from the middle and 
lower bands arising from the Wick decomposition. The effective chemical potential has three contributions that
are all linear in $U$, and their
sum is negative for repulsive interactions in our parameter range
$J=1$ and $V=0.2$--$0.6$. In the particle--hole-symmetric,
single-orbital $t$--$V$ model at half filling, the corresponding
contributions cancel. Such a cancellation is not enforced in our
multiorbital model: the interorbital hopping $J$, together with the
intrachain hopping $t$, forms triangular hopping paths that break
the physical particle--hole symmetry. The effective
chemical-potential shift favors hole doping  and can promote incommensurate density
correlations. We can also 
absorb the $\partial_x\phi$ term into the Gaussian Hamiltonian by shifting
$\phi$ to introduce a spatial mismatch into the Umklapp term.
The $\phi$ sector of Eq.~\eqref{lowenergy} becomes
\begin{equation}
H_\phi
=\int_0^\ell dx\left[
\frac{v}{2\pi K}(\partial_x\widetilde\phi)^2
-\lambda_u\cos(4\widetilde\phi-Q_{\mathrm{mis}}x)
\right]
\end{equation}

where
$\widetilde\phi(x)=\phi(x)+K\mu_{\mathrm{eff}}x/v$ and
$Q_{\mathrm{mis}}=4K\mu_{\mathrm{eff}}/v$.
In the soliton description, the gradient favors $\partial_x\widetilde\phi=0$, while locking by the cosine requires its argument
$4\widetilde\phi-Q_{\mathrm{mis}}x$ to remain at a minimum
of the cosine potential, $ie$. $\partial_x\widetilde\phi=K\mu_{\mathrm{eff}}/v$.
Remaining on the same branch of minima of the cosine potential
incurs a gradient-energy cost, whereas a transition to an adjacent
branch introduces a soliton. Therefore a finite soliton density can move the
density modulation away from commensurability. Provided that
the pairing perturbation does not gap the resulting density
mode, this mechanism supports an ICDW
regime.

We analyze the competing perturbations using a second-order
renormalization-group (RG) expansion. The RG flow equations [Eq.~\eqref{eq:app-RG-full}] are
\begin{equation}
\begin{aligned}
\frac{dK}{dl}
&=\frac{\delta^2}{2}
-2K^2y^2J_0(\zeta), \quad
\frac{d\delta}{dl}
=\left(2-\frac1K\right)\delta,\\
\frac{dy}{dl}
&=(2-4K)y, \quad
\frac{d\zeta}{dl}
=\zeta-4Ky^2J_1(\zeta),
\end{aligned}
\end{equation}
where
$y=\frac{g_u}{\pi v},
\delta=\frac{4\alpha\Delta}{v},
\zeta=Q_{\mathrm{mis}}\alpha_0.$
Motivated by the observed exponential decay of connected
pairing correlations and the gapless character of the system, we consider the RG flow with $\delta=0$, treating 
pairing as irrelevant. There are two fixed points of the flow. 
$(K_*,y_*,\zeta_*)=(1/2,0,0)$ is unstable, since any initial incommensurability cannot approach
this point, while $\left(1/2,-1.52,-2.40\right)$ is stable under a small deviation
from the fixed point. Applying the fourth-order Runge–Kutta method to
solve the coupled differential equations from $\left(1/2,-1.52,-2.40\right)$ towards
$(K_c(U), y_c(U),\zeta_c(U))$, we obtain the phase transition point
at $U_c\simeq0.27,K_{\mathrm{eff},c}\simeq0.47$.

Within the proposed low-energy description, the TSC phase is a
gapped superconducting regime in which pairing locks the dual
field and supports localized Majorana boundary modes. Upon
crossing the TSC--IC transition~\cite{gangadharaiahMajoranaEdgeStates2011a}, the system enters a gapless
incommensurate regime, and the bulk gap protecting the Majorana
modes is lost. The effective Luttinger
parameter decreases rapidly from the TSC--IC boundary, where fitting in the 
gapped regime fails, to approximately $1/2$ near the
numerically identified IC--C boundary, consistent with our prediction
of the effective Luttinger parameter (Fig.~\ref{fig:3}(f)).
The Luttinger parameter can be obtained from density correlations and density and
bond oscillations under OBCs (Sec.~\ref{app:luttinger_fits} of the Supplemental Material). We apply a pure Tomonaga--Luttinger theory since Umklapp
and pairing are irrelevant in the ICDW regime to acquire $K_{\mathrm{eff}}$.

Within our truncated RG equations,
the dilute-soliton IC--C point with $(y_*,\zeta_*)=(0,0)$ cannot be
approached from nonzero incommensurability, where doping appears
just to the left of the transition, creating dilute solitons, whereas the Umklapp 
 coupling $y$ appears just to its right. The dilute-soliton IC--C point
 is characterized by continuously changing doping. The negative fixed point
in Eqs.~\eqref{eq:app-IC-RG-critical-estimate} and~\eqref{eq:app-fixedpoints} instead
provides a competition between Umklapp and incommensurability
couplings. Both doping and Umklapp scattering remain 
strong throughout the parameter regime of interest. The IC-C fixed point
is associated with a discontinuous change in doping, as shown in Fig.~\ref{fig:3}(c).
In the ICDW phase, doping dominates, resulting in 
 a finite deviation from half filling. In the CDW phase, when $K<1/2$, 
 Umklapp is relevant and doping is irrelevant.
The Umklapp process locks the system at half filling. 

\par\smallskip\noindent\textit{Regime III:  CDW regime.}---
At strong coupling, 
the system exhibits  period-two density modulations
on all three orbitals (Fig.~\ref{fig:S4}(a,b)). The average filling remains approximately
one half, while the connected density correlations decay
approximately exponentially, indicating
short-ranged fluctuations around the ordered density pattern.
These observations characterize a strongly ordered $ABC$-CDW
state close to a classical alternating charge configuration.

\par\smallskip\noindent\textit{Discussion}---We have investigated interacting
Lieb-Kitaev chain model.
Our results distinguish two aspects of Majorana stability:
finite-size protection within the topological phase and the
survival of the phase itself. The interaction and geometry-induced
chemical-potential shift offers a microscopic explanation
for the departure from half filling and the discontinuous change in doping. We are unable to derive the TSC--IC transition point because
its determination may require a higher-order RG analysis.
Comparing these energy branches
and compensating the filling shift with an external chemical
potential would help emphasize the role of multiorbital flat-band systems. Such tests would clarify how quantum geometry,
density feedback, and commensurate locking jointly shape
the phase diagram.


\par\smallskip\noindent\textit{Acknowledgments}---JX and XYX are supported by the National Natural Science Foundation of China (Grant Nos. 12447103, 
No. 12674187), the National Key R\&D Program of China (Grant No. 2022YFA1402702), 
Yangyang Development Fund, and Shanghai Jiao Tong University 2030 Initiative. ZXY and KTL are supported 
by the Ministry of Science and Technology, China, The New Cornerstone Foundation, and the Hong Kong Research Grants
Council through Grant Nos. MOST23SC01-A, No.
RFS2021-6S03, No. C6053-23G, No. AoE/P-701/20,
AoE/P-604/25R, No. 16309223, No. 16311424 and No.
16300325. The computations in this paper were run on the 
Siyuan-1 and $\pi$ 2.0 clusters supported by the Center for 
High Performance Computing at Shanghai Jiao Tong University.

\bibliographystyle{apsrev4-2}
\bibliography{refs}

\clearpage
\onecolumngrid
\makeatletter
\let\@makecaption\PRLoriginalcaption
\makeatother
\normalsize
\setlength{\jot}{3pt}

\begin{center}
    {\Large\bfseries Supplementary Material for ``Quantum Metric Control of Topological Superconductivity and Incommensurate Charge Density Wave in an Interacting Lieb-Kitaev Chain Model''}

    \vspace{0.5em}

\end{center}

\setcounter{secnumdepth}{2}
\setcounter{subsection}{0}
\renewcommand{\thesubsection}{\Alph{subsection}}
\setcounter{section}{0}
\setcounter{equation}{0}
\setcounter{figure}{0}
\setcounter{table}{0}

\renewcommand{\thesection}{S\arabic{section}}
\renewcommand{\theequation}{S\arabic{equation}}
\renewcommand{\thefigure}{S\arabic{figure}}
\renewcommand{\thetable}{S\arabic{table}}
\renewcommand{\theHequation}{S\arabic{equation}}
\renewcommand{\theHfigure}{S\arabic{figure}}
\renewcommand{\theHtable}{S\arabic{table}}

\section{Low-energy effective theory}
\label{app:lowenergy}

The Lieb--Kitaev model has three orbitals per unit cell, with its
quadratic Hamiltonian written in Nambu representation to incorporate hopping and
pairing. Its normal-state spectrum features a nearly flat middle band;
in the regime where its bandwidth is smaller than the interaction scale
and the interaction scale is smaller than the interband gap, a single-band projection
provides a natural starting point. We project the quadratic Hamiltonian
onto the middle-band particle and hole sectors and use the same
orbital-to-band transformation to express the interaction in this basis.
In the continuum limit, linearizing the normal middle-band dispersion
about the two Fermi points yields a Tomonaga--Luttinger liquid 
theory in terms of the density field $\phi$ and its dual field $\theta$,
while pairing generates a $\cos(2\theta)$ perturbation. Terms in the projected
interaction renormalize the Luttinger parameter, induce an
effective chemical-potential shift, and allow Umklapp scattering at
half filling of the effective band, represented by $\cos(4\phi)$.

The second-order RG analysis shows a sequence of phase transitions from a topological
superconducting (TSC) phase through an incommensurate charge-density-wave
(ICDW) regime to a commensurate CDW phase. In the TSC phase, 
the weak-coupling description breaks down and the numerical
Luttinger parameter $K_{\mathrm{eff}}$ approaches infinity. Within the ICDW regime, $K_{\mathrm{eff}}$ 
decreases with increasing interaction. The low-energy excitations 
are governed by solitons of the ICDW order, which induce short-range 
superconducting correlations. Eventually, $K_{\mathrm{eff}}$ goes below $1/2$ to an Umklapp-relevant
regime. The interval $U_c\simeq0.25$--$0.26$ is therefore
considered an IC--C boundary close to our DMRG results.

\subsection{Microscopic Hamiltonian and noninteracting projection}

Let $j=1,\ldots,L$ label unit cells, $a$ be the cell spacing, and
$\alpha\in\{A,C,B\}$ label orbitals. The total number
of microscopic sites is $N=3L$.  The onsite energies are
$(v_A,v_C,v_B)=(V,-V,0)$. Although the finite chain has open boundaries, its bulk low-energy
theory can be constructed using the translationally invariant reference
system. The bulk Hamiltonian of the interacting Lieb--Kitaev model is
\begin{align}
H&=H_0+H_\Delta+H_U,
\label{eq:app-micro}\\
H_0&=-t\sum_{j,\alpha}
\bigl(c^\dagger_{j,\alpha}c_{j+1,\alpha}+\mathrm{H.c.}\bigr)
+V\sum_j(n_{j,A}-n_{j,C})
\nonumber\\
&\quad-J\sum_j\Bigl[
(c^\dagger_{j,A}+c^\dagger_{j,C})c_{j,B}
+(c^\dagger_{j+1,A}+c^\dagger_{j+1,C})c_{j,B}
+\mathrm{H.c.}\Bigr],\\
H_\Delta&=\Delta\sum_{j,\alpha}
\bigl(c^\dagger_{j+1,\alpha}c^\dagger_{j,\alpha}
+\mathrm{H.c.}\bigr),\\
\label{pair}
H_U&=U\sum_{j,\alpha}
\left(n_{j,\alpha}-\frac12\right)
\left(n_{j+1,\alpha}-\frac12\right).
\end{align}

We first project the noninteracting Hamiltonian onto the flat-band basis derived in Ref.~\cite{huAnomalousCoherenceLength2025}.  In the basis
$\bm c_q=(c_{q,A},c_{q,C},c_{q,B})^T$,
\begin{align}
h_0(q)&=
\begin{pmatrix}
-2t\cos q-\mu+V & 0 & f(q)\\
0 & -2t\cos q-\mu-V & f(q)\\
f(q)^* & f(q)^* & -2t\cos q-\mu
\end{pmatrix},\\
\epsilon_0(q)&=-2t\cos q-\mu,\qquad
\epsilon_\pm(q)=-2t\cos q-\mu
\pm\sqrt{V^2+4J^2(1+\cos q)}.
\end{align}

Here $f(q)=-J(1+e^{-iq})$ is the interorbital hopping amplitude.
Let $\mathcal E(q)=\sqrt{V^2+2|f(q)|^2}$ and label the middle, upper, and lower energy 
bands by $(0,\pm)$. For $V>0$, a normalized middle-band eigenvector is
\begin{equation}
|u_0(q)\rangle=\frac1{\mathcal E(q)}
\begin{pmatrix}-f(q)\\f(q)\\V\end{pmatrix}.
\label{eq:app-u0}
\end{equation}

At the right Fermi point $k_F=\pi/2a$, the orbital components are
\begin{equation}
u_{A0}=\frac{J(1-i)}{\sqrt{V^2+4J^2}},
\qquad
u_{B0}=\frac{V}{\sqrt{V^2+4J^2}},
\qquad
u_{C0}=-\frac{J(1-i)}{\sqrt{V^2+4J^2}}.
\label{eq:app-u0-Fermi}
\end{equation}

The components at the left Fermi point satisfy
$u_{\alpha0}(-q_F)=u_{\alpha0}(q_F)^*$. A normalized lower-band eigenvector, with eigenvalue
$\epsilon_-(q)=\epsilon(q)-\mathcal E(q)$, is
\begin{equation}
|u_-(q)\rangle
=\frac{1}{2\mathcal E(q)}
\begin{pmatrix}
\mathcal E(q)-V\\
\mathcal E(q)+V\\
-2f(q)^*
\end{pmatrix},
\label{eq:app-u-minus}
\end{equation}

\begin{equation}
u_{A-}
=\frac{\sqrt{V^2+4J^2}-V}
{2\sqrt{V^2+4J^2}},\qquad
u_{B-}
=\frac{J(1+i)}{\sqrt{V^2+4J^2}},\qquad
u_{C-}
=\frac{\sqrt{V^2+4J^2}+V}
{2\sqrt{V^2+4J^2}}.
\label{eq:app-u-minus-Fermi}
\end{equation}

The quantum metric is
\begin{equation}
g_{kk}(k)
=
\langle\partial_k u_0|
\left(1-|u_0\rangle\langle u_0|\right)
|\partial_k u_0\rangle
=
\frac{2J^2V^2}{
\left\{V^2+4J^2[1+\cos(ka)]\right\}^2}.
\end{equation}

Its Brillouin-zone integral defines the QML
\begin{equation}
\xi_{\mathrm{QML}}^{(0)}(V,J)
=
\int_{-\pi}^{\pi}\frac{dk}{2\pi}\,g_{kk}(k)
=
\frac{2J^2(V^2+4J^2)}
{V(V^2+8J^2)^{3/2}}.
\label{eq:IQML-bare}
\end{equation}

Choosing an orthonormal eigenvector from each projector defines
\begin{equation}
d_{q,s}=\sum_\alpha u_{\alpha s}(q)^*c_{q,\alpha},\qquad
c_{q,\alpha}=\sum_{s=0,\pm}u_{\alpha s}(q)d_{q,s},
\label{trans}
\end{equation}

and $H_0=\sum_{q,s}\epsilon_s(q)d^\dagger_{q,s}d_{q,s}$,
up to a constant. The middle band is exactly flat only when $t=0$; the
nonzero $t$ used here gives it bandwidth $4t$. 
Projecting onto the middle band and defining $d_q\equiv d_{q,0}$,
we obtain
\begin{equation}
H_0^{\mathrm{proj}}
=\sum_q(-2t\cos q)d_q^\dagger d_q.
\end{equation}

The middle band is half filled. In one-dimensional systems, the low-energy excitations
can be expanded at two Fermi points, $k_F=\pm\pi/2a$. That gives
$\epsilon_0(\pm\pi/2+ap)\simeq\pm v_Fp$, with $v_F=2ta$. Below we set $a=1$.
Following the standard bosonization procedure~\cite{giamarchi2004}, we linearize the
projected dispersion around the two Fermi points and obtain
\begin{equation}
H_{\mathrm{LL}}^{(0)}
=\frac{v_F}{2\pi}\int_0^\ell dx\,
\left[(\partial_x\theta)^2+(\partial_x\phi)^2\right],
\qquad v_F=2ta.
\label{noninteract}
\end{equation}

Here $\ell$ is the continuum length of the open chain, $\phi$ describes
the smooth density fluctuation through
$\delta\rho=-\partial_x\phi/\pi$, and $\theta$ is its dual phase field.
This is the noninteracting Tomonaga--Luttinger Hamiltonian with $K=1$.
The bare chemical potential remains zero; any effective
chemical-potential shift introduced below arises from the projected
interaction.

\subsection{Projected interaction}

The lattice density interaction gives, in a periodic bulk,
\begin{equation}
H_U=\frac1{2L}\sum_{k,p,q,\alpha}2U\cos q\,
c^\dagger_{k+q,\alpha}c^\dagger_{p-q,\alpha}c_{p,\alpha}c_{k,\alpha}
-U\sum_{k,\alpha}c^\dagger_{k,\alpha}c_{k,\alpha}+\mathrm{const}.
\label{interaction}
\end{equation}

After projection, it yields:
\begin{equation}
\begin{aligned}
H_U^{(4)}
={}&\frac{U}{L}
\sum_{\alpha=A,B,C}
\sum_{k_1,k_2,k_3,k_4}
\sum_{s_1,s_2,s_3,s_4\in\{0,+,-\}}
\delta_{k_1+k_2,k_3+k_4}\,
e^{i(k_3-k_2)}
\\
&\times
u_{\alpha s_1}^{*}(k_1)
u_{\alpha s_2}^{*}(k_2)
u_{\alpha s_3}(k_3)
u_{\alpha s_4}(k_4)\,
d_{k_1,s_1}^{\dagger}
d_{k_2,s_2}^{\dagger}
d_{k_3,s_3}
d_{k_4,s_4}.
\end{aligned}
\label{eq:app-projected-interaction-momentum}
\end{equation}

where $\alpha$ labels orbitals and $s$ labels bands, with $(0,+,-)$ denoting the middle, upper, and 
lower bands, respectively. After projecting the interaction onto the middle band, we normal order
the resulting four-fermion operators with respect to the noninteracting
half-filled Fermi sea. Wick's theorem separates the interaction into
a residual normal-ordered four-fermion term, a bilinear contribution
from direct and exchange contractions, and a constant.

In the noninteracting half-filled Fermi sea, we have
$\langle d_k^\dagger d_p\rangle_0=\delta_{kp}n_k^{(0)}$
with $n_k^{(0)}=\Theta[-\epsilon_0(k)]$.
Wick's theorem gives:
\begin{align}
d^\dagger_{k+q}d^\dagger_{p-q}d_p d_k
={}&:d^\dagger_{k+q}d^\dagger_{p-q}d_p d_k:_0
\nonumber\\
&+\delta_{q,0}
\left[
n_k^{(0)}:d_p^\dagger d_p:_0
+n_p^{(0)}:d_k^\dagger d_k:_0
\right]
\nonumber\\
&-\delta_{q,p-k}
\left[
n_p^{(0)}:d_k^\dagger d_k:_0
+n_k^{(0)}:d_p^\dagger d_p:_0
\right]
\nonumber\\
&+\left(\delta_{q,0}-\delta_{q,p-k}\right)
n_k^{(0)}n_p^{(0)}.
\label{eq:app-Wick}
\end{align}

Here $:\cdots:_0$ denotes normal ordering relative to this reference
state. The single contractions generate direct and exchange
bilinears, whereas the double contractions contribute a constant.

The projected interaction vertex is defined as
\begin{equation}
\Gamma(k,p;q)
=2U\cos q\sum_{\alpha=A,B,C}
u_{\alpha0}^{*}(k+q)\,
u_{\alpha0}^{*}(p-q)\,
u_{\alpha0}(p)\,
u_{\alpha0}(k).
\label{eq:app-vertex}
\end{equation}

Here $k$ and $p$ are the incoming momenta, $q$ is the momentum transfer,
and $u_{\alpha0}(k)$ is the orbital-$\alpha$ component of the
middle-band Bloch eigenvector. Substituting Eq.~\eqref{eq:app-Wick} into the projected interaction Eq.~\eqref{eq:app-projected-interaction-momentum} yields
\begin{align}
H_U^{(4)}
={}&\frac{1}{2L}\sum_{k,p,q}\Gamma(k,p;q)
:d^\dagger_{k+q}d^\dagger_{p-q}d_p d_k:_0
\nonumber\\
&+\sum_k\left[\Sigma_{\mathrm{mid}}(k)+\Sigma_{\mathrm{bg}}(k)\right]:d_k^\dagger d_k:_0
+const,
\label{eq:app-normal-ordered-interaction}\\
\Sigma_{\mathrm{mid}}(k)
={}&\frac{1}{L}\sum_p
\left[\Gamma(k,p;0)-\Gamma(k,p;p-k)\right]n_p^{(0)}
\nonumber\\
={}&\frac{2U}{L}\sum_{p,\alpha}
\left[1-\cos(p-k)\right]
|u_{\alpha0}(k)|^2|u_{\alpha0}(p)|^2 n_p^{(0)},
\label{eq:app-mid-selfenergy}\\
\Sigma_{\mathrm{bg}}(k)
={}&\frac{2U}{L}\sum_{p,\alpha}
\left[1-\cos(p-k)\right]
|u_{\alpha0}(k)|^2|u_{\alpha-}(p)|^2.
\label{eq:app-bg-selfenergy} 
\end{align}

To construct the low-energy Hamiltonian, we retain only the
middle-band operators after Wick contraction and projection.
However, the Bloch amplitudes of the eliminated bands remain
in the coefficients generated by their contractions.
For the band-index configuration in Eq.~\eqref{eq:app-projected-interaction-momentum} with $(s_1,s_2,s_3,s_4)=(0,0,0,0)$, Wick decomposition produces
a normal-ordered four-fermion interaction, the middle-band
bilinear correction $\Sigma_{\mathrm{mid}}$, and a constant.
The configurations $(-,0,0,-)$ and $(0,-,-,0)$ generate the
direct lower-band contractions, whereas $(-,0,-,0)$ and
$(0,-,0,-)$ generate the exchange contractions.
Since the lower band is fully occupied,
$\langle d^\dagger_{k,-}d_{p,-}\rangle_0=\delta_{kp}$,
these contractions together  give $\Sigma_{\mathrm{bg}}$.
The upper-band contractions vanish because
$\langle d^\dagger_{k,+}d_{p,+}\rangle_0=0$.
Finally, the configuration $(-,-,-,-)$ contributes only a constant
after projection onto the filled lower band.

\subsection{Bilinear contribution: chemical-potential and velocity shifts}

We now consider the bilinear term generated from Eqs.~\eqref{eq:app-normal-ordered-interaction}, \eqref{eq:app-mid-selfenergy},
and~\eqref{eq:app-bg-selfenergy}. First, we evaluate Eq.~\eqref{eq:app-mid-selfenergy}. 
To obtain its leading low-energy contribution, we expand it in a
small momentum displacement from each Fermi point. Its constant part
shifts the energies of both branches equally and therefore acts as an
effective chemical potential, whereas its linear part modifies the
branch velocities. The expansion is
\begin{equation}
\Sigma_{\mathrm{mid}}(\pm k_F+ap)
=\Sigma_{\mathrm{mid},F}
\pm  p\,\Sigma_{\mathrm{mid},F}'
+O((p)^2),
\end{equation}

where $\Sigma_{\mathrm{mid}}(k)$ is an even function of $k$ and
$\Sigma_{\mathrm{mid},F}=\Sigma_{\mathrm{mid}}(k_F)$ and
$\Sigma_{\mathrm{mid},F}'=
\partial_k\Sigma_{\mathrm{mid}}(k)|_{k_F}$.
We start from the bilinear generated by normal ordering, and expand the excitation
at $k=k_{F}+p$. Here $a$ is the unit-cell spacing set to 1, $p$ is the wave-vector displacement
from either Fermi point, and $\Lambda$ is the corresponding cutoff,
chosen such that $a\Lambda\ll1$. Defining $R_p\equiv d_{k_F+p}$ and
$L_p\equiv d_{-k_F+p}$, we have
\begin{equation}
\begin{aligned}
H_{\mathrm{bil}}^{\mathrm{mid}}
&=\sum_k\Sigma_{\mathrm{mid}}(k):d_k^\dagger d_k:_0\\
&\simeq\sum_{|p|<\Lambda}
\left[
\Sigma_{\mathrm{mid}}(k_F+p):R_p^\dagger R_p:_0
+\Sigma_{\mathrm{mid}}(-k_F+p):L_p^\dagger L_p:_0
\right]\\
&\simeq
\Sigma_{\mathrm{mid},F}\sum_{|p|<\Lambda}
\left[:R_p^\dagger R_p:_0+:L_p^\dagger L_p:_0\right]\\
&\quad
+\Sigma_{\mathrm{mid},F}'\sum_{|p|<\Lambda}p
\left[:R_p^\dagger R_p:_0-:L_p^\dagger L_p:_0\right].
\end{aligned}
\label{eq:app-bilinear-expanded}
\end{equation}

To obtain the local continuum Hamiltonian, we introduce the
slowly varying fields in the bulk reference system, with the properties below.
\begin{equation}
\begin{gathered}
R(x)=\frac{1}{\sqrt{\ell}}\sum_p e^{ipx}R_p,
\qquad
L(x)=\frac{1}{\sqrt{\ell}}\sum_p e^{ipx}L_p,\\
\begin{aligned}
\sum_p:R_p^\dagger R_p:_0
&=\int dx\,:R^\dagger(x)R(x):_0
=\int dx\,\rho_R(x),\\
\sum_p:L_p^\dagger L_p:_0
&=\int dx\,:L^\dagger(x)L(x):_0
=\int dx\,\rho_L(x),\\
\sum_p p:R_p^\dagger R_p:_0
&=-i\int dx\,:R^\dagger(x)\partial_xR(x):_0,\\
\sum_p p:L_p^\dagger L_p:_0
&=-i\int dx\,:L^\dagger(x)\partial_xL(x):_0.
\end{aligned}
\end{gathered}
\label{eq:app-chiral-fourier-identities}
\end{equation}

Thus, the constant part couples to the total smooth density,
whereas the linear part has the same operator structure as the
chiral kinetic energy. On the finite interval, with open-boundary
conditions imposed on the continuum fields, we obtain
\begin{align}
H_{\mathrm{bil}}^{\mathrm{mid}}
\simeq{}&
\Sigma_{\mathrm{mid},F}
\int_0^\ell dx\,(\rho_R+\rho_L)
\nonumber\\
&-i\Sigma_{\mathrm{mid},F}'\int_0^\ell dx\,
\left[:R^\dagger\partial_xR:_0
-:L^\dagger\partial_xL:_0\right]
\label{midbil}
\end{align}

The effective chemical-potential contribution is therefore
$\mu_{\mathrm{mid}}=-\Sigma_{\mathrm{mid},F}$, while
$\Sigma_{\mathrm{mid},F}'$ is the interaction-induced velocity correction.
Within the standard bosonization procedure, the total density
$\rho_R+\rho_L$ in the first term is replaced by
$-\partial_x\phi/\pi$.
\begin{equation}
\mu_{\mathrm{mid}}=-\Sigma_{\mathrm{mid},F},
\qquad
H_{\mu,\mathrm{mid}}
=-\mu_{\mathrm{mid}}\int_0^\ell dx\,(\rho_R+\rho_L)
=\frac{\mu_{\mathrm{mid}}}{\pi}
\int_0^\ell dx\,\partial_x\phi.
\label{eq:app-mu-middle}
\end{equation}

For the second term, using
$\rho_{R,L}=(-\partial_x\phi\pm\partial_x\theta)/(2\pi)$,
this identity gives
\begin{equation}
\begin{aligned}
H_{g_4,\mathrm{mid}}
&=-i\delta v_{\mathrm{mid}}\int_0^\ell dx\,
\left[:R^\dagger\partial_xR:_0
-:L^\dagger\partial_xL:_0\right]\\
&=\frac{g_{4,\mathrm{mid}}}{4\pi^2}
\int_0^\ell dx\,
\left[(\partial_x\phi)^2+(\partial_x\theta)^2\right],
\end{aligned}
\label{eq:app-g4-bosonized}
\end{equation}

where
$g_{4,\mathrm{mid}}=2\pi \,\Sigma_{\mathrm{mid},F}'$
denotes the contribution from the interaction-induced velocity correction. Now 
we express these terms in terms of the parameters $V,J$. We define
\begin{equation}
n_\alpha^{(0)}
=\frac{1}{L}\sum_p
|u_{\alpha0}(p)|^2n_p^{(0)},
\qquad
\chi_\alpha^{(0)}
=\frac{1}{L}\sum_p
\cos p\,|u_{\alpha0}(p)|^2n_p^{(0)}.
\label{n,chi}
\end{equation}

Equation~\eqref{eq:app-mid-selfenergy} yields

\begin{equation}
\Sigma_{\mathrm{mid}}(k)
=2U\sum_\alpha |u_{\alpha0}(k)|^2
\left[n_\alpha^{(0)}-\chi_\alpha^{(0)}\cos k\right].
\label{eq:app-Sigma-for-matching}
\end{equation}

At $k_F=\pi/2$, we obtain
\begin{align}
\mu_{\mathrm{mid}}
&=-2U\sum_\alpha
|u_{\alpha0}(k_F)|^2n_\alpha^{(0)},\\
g_{4,\mathrm{mid}}
&=4\pi U\sum_\alpha
\left[
\left.\partial_k|u_{\alpha0}(k)|^2\right|_{k=k_F}
n_\alpha^{(0)}
+|u_{\alpha0}(k_F)|^2\chi_\alpha^{(0)}
\right].
\label{eq:app-mu-g4-orbital}
\end{align}

To evaluate these expressions, we use the middle-band orbital weights from~\eqref{eq:app-u0-Fermi}
\begin{align}
|u_{A0}(k_F)|^2=|u_{C0}(k_F)|^2
&=\frac{2J^2}{V^2+4J^2},
&
|u_{B0}(k_F)|^2
&=\frac{V^2}{V^2+4J^2},\\
\left.\partial_k|u_{A0}(k)|^2\right|_{k=k_F}
=\left.\partial_q|u_{C0}(k)|^2\right|_{k=k_F}
&=-\frac{2J^2V^2}{(V^2+4J^2)^2},
&
\left.\partial_k|u_{B0}(k)|^2\right|_{k=k_F}
&=\frac{4J^2V^2}{(V^2+4J^2)^2}.
\end{align}

Substituting these equations into~\eqref{n,chi}, we have
\begin{align}
n_B^{(0)}
&=\frac{1}{2\pi}\int_{-\pi/2}^{\pi/2}
dk\,\frac{V^2}{V^2+4J^2(1+\cos k)}
\nonumber\\
&=\frac{2V}{\pi\sqrt{V^2+8J^2}}
\arctan\!\left(\frac{V}{\sqrt{V^2+8J^2}}\right),
\label{eq:app-N-function}\\
\chi_B^{(0)}
&=\frac{1}{2\pi}\int_{-\pi/2}^{\pi/2}
dk\,\frac{V^2\cos k}{V^2+4J^2(1+\cos k)}
\nonumber\\
&=\frac{V^2}{8J^2}
-\frac{V^2+4J^2}{4J^2}n_B^{(0)},
\label{eq:app-C-function}\\
n_A^{(0)}=n_C^{(0)}
&=\frac14-\frac{n_B^{(0)}}{2},
\qquad
\chi_A^{(0)}=\chi_C^{(0)}
=\frac{1}{2\pi}-\frac{\chi_B^{(0)}}{2}.
\label{eq:app-AC-contractions}
\end{align}

The resulting parameter expressions are
\begin{align}
\mu_{\mathrm{mid}}
&=-\frac{2U}{V^2+4J^2}
\left[J^2+(V^2-2J^2) n_B^{(0)}\right],
\label{eq:app-mu-parameters}\\
g_{4,\mathrm{mid}}
&=4\pi U\left[
\frac{4J^2V^2}{(V^2+4J^2)^2}
\left(\frac32n_B^{(0)}-\frac14\right)
+\frac{2J^2}{\pi(V^2+4J^2)}
+\frac{V^2-2J^2}{V^2+4J^2}\chi_B^{(0)}
\right].
\label{eq:app-g4-parameters}
\end{align}

Then we calculate~\eqref{eq:app-bg-selfenergy}. From~\eqref{midbil}, we just need to
calculate $\Sigma_{\mathrm{bg},F}$ and $\Sigma_{\mathrm{bg},F}'$. Similarly, we 
define
\begin{equation}
n_\alpha^-
=\frac{1}{L}\sum_p|u_{\alpha-}(p)|^2, \qquad
\chi_\alpha^-
=\frac{1}{L}\sum_p\cos p\,|u_{\alpha-}(p)|^2,
\label{eq:app-lower-band-contractions}
\end{equation}

and using the evenness of the orbital weights in momentum,
Eq.~\eqref{eq:app-bg-selfenergy} becomes
\begin{equation}
\Sigma_{\mathrm{bg}}(k)
=2U\sum_\alpha|u_{\alpha0}(k)|^2
\left(n_\alpha^--\chi_\alpha^-\cos k\right).
\label{eq:app-bg-selfenergy-orbital}
\end{equation}

Since the lower band is completely filled, its contribution on
the $B$ orbital can be evaluated over the full Brillouin zone from~\eqref{eq:app-u-minus}
and~\eqref{eq:app-u-minus-Fermi}
\begin{align}
n_B^-
&=\frac{1}{2\pi}\int_{-\pi}^{\pi}
dp\,\frac{2J^2(1+\cos p)}
{V^2+4J^2(1+\cos p)}
\nonumber\\
&=\frac12\left(1-\frac{V}{\sqrt{V^2+8J^2}}\right),
\label{eq:app-bg-nB}\\
\chi_B^-
&=\frac{1}{2\pi}\int_{-\pi}^{\pi}
dp\,\frac{2J^2(1+\cos p)\cos p}
{V^2+4J^2(1+\cos p)}
\nonumber\\
&=-\frac{V^2}{8J^2}
+\frac{V^2+4J^2}{8J^2}
\frac{V}{\sqrt{V^2+8J^2}},
\label{eq:app-bg-chiB}
\end{align}

where $V>0$ and $J\neq0$.
Normalization of the lower-band eigenvector then gives $|u_{A0}(k)|^2=|u_{C0}(k)|^2$ and
\begin{equation}
n_A^-+n_C^-=1-n_B^-,
\qquad
\chi_A^-+\chi_C^-=-\chi_B^-.
\label{eq:app-bg-AC-contractions}
\end{equation}

At the  Fermi point $k_F=\pi/2$, the self-energy
and its derivative  are
\begin{align}
\Sigma_{\mathrm{bg},F}
&=\frac{U}{V^2+4J^2}
\left[4J^2+2(V^2-2J^2)n_B^-\right],
\label{eq:app-bg-selfenergy-F}\\
\Sigma_{\mathrm{bg},F}'
&=U\left[
\frac{4J^2V^2}{(V^2+4J^2)^2}(3n_B^--1)
+\frac{2(V^2-2J^2)}{V^2+4J^2}\chi_B^-
\right].
\label{eq:app-bg-selfenergy-derivative}
\end{align}

Therefore, these two terms also shift the chemical potential and effective velocity
$\mu_{\mathrm{bg}}=-\Sigma_{\mathrm{bg},F}$ and
$g_{4,\mathrm{bg}}=2\pi \Sigma_{\mathrm{bg},F}'$.

The third term comes from Eq.~\eqref{interaction} as $-U\sum_{k,\alpha}c^\dagger_{k,\alpha}c_{k,\alpha}$.
Using the unitarity of the orbital-to-band transformation,
this bilinear term becomes
\begin{equation}
-U\sum_{k,\alpha}c^\dagger_{k,\alpha}c_{k,\alpha}
=-U\sum_{k,s}d^\dagger_{k,s}d_{k,s}.
\end{equation}

After projection onto the middle band, the filled lower band
contributes only a constant, leaving
\begin{equation}
\begin{aligned}
H_{\mathrm{cent}}^{\mathrm{proj}}
&=-U\sum_k:d_k^\dagger d_k:_0+\mathrm{const}\\
&\simeq\frac{U}{\pi}\int_0^\ell dx\,\partial_x\phi+\mathrm{const}.
\label{third}
\end{aligned}
\end{equation}

Thus, this term contributes a chemical-potential shift
$\mu_{\mathrm{cent}}=U$.  Combining the results of Eqs.~\eqref{eq:app-mu-parameters}, \eqref{eq:app-g4-parameters}, \eqref{eq:app-bg-selfenergy-F}, \eqref{eq:app-bg-selfenergy-derivative}, and~\eqref{third},
we have 
\begin{equation}
H_{\mathrm{bil}}
=\frac{\mu_{\mathrm{eff}}}{\pi}
\int_0^\ell dx\,\partial_x\phi
+\frac{g_4}{4\pi^2}
\int_0^\ell dx\,
\left[(\partial_x\phi)^2+(\partial_x\theta)^2\right],
\label{eq:app-bilinear-final}
\end{equation}

where
\begin{equation}
\begin{aligned}
\mu_{\mathrm{eff}}
&=U(1-\Sigma_{\mathrm{mid},F}-\Sigma_{\mathrm{bg},F})\\
g_4
&=2\pi \left[
\Sigma_{\mathrm{mid},F}'+\Sigma_{\mathrm{bg},F}'
\right]
\end{aligned}
\label{eq:app-bilinear-coefficients}
\end{equation}

Here we factor out \(U\) by $U\Sigma \rightarrow \Sigma$. The effective chemical potential has three contributions that
are all linear in $U$, and their
sum is negative for repulsive interactions in our parameter range
$J=1$ and $V=0.2$--$0.6$. In the particle--hole-symmetric,
single-orbital $t$--$V$ model at half filling, the corresponding
contributions cancel. Such a cancellation is not enforced in our
multiorbital model: the interorbital hopping $J$, together with the
intrachain hopping $t$, forms triangular hopping paths that break
the physical particle--hole symmetry. The effective
chemical-potential shift favors hole doping  and can promote incommensurate density
correlations.

\subsection{Four-fermion scattering processes}
\label{sec:app-four-fermion}

After extracting the bilinear contributions, the remaining
normal-ordered interaction is
\begin{equation}
H_{U}^{(4)}
=\frac{1}{2L}\sum_{k,p,q}\Gamma(k,p;q)
:d^\dagger_{k+q}d^\dagger_{p-q}d_p d_k:_0,
\label{eq:app-residual-quartic}
\end{equation}
\begin{equation}
\Gamma(k,p;q)
=2U\cos q\sum_\alpha
u_{\alpha0}^{*}(k+q)u_{\alpha0}^{*}(p-q)
u_{\alpha0}(p)u_{\alpha0}(k).
\end{equation}

We expand the momentum dependence around the Fermi points
$k_F=\pi/2$, and retain the leading
low-energy contributions. These comprise direct interbranch
scattering, exchange interbranch scattering, and the
commensurate Umklapp process. 

\paragraph{Direct interbranch scattering.}

For incoming particles near $k=k_F$ and $p=-k_F$, a small
momentum transfer $q\simeq0$ leaves both particles on their
original branches. The corresponding vertex is
\begin{equation}
\Gamma_{\mathrm D}
=\Gamma(k_F,-k_F;0)=2U\sum_\alpha
|u_{\alpha0}(k_F)|^2|u_{\alpha0}(-k_F)|^2.
\end{equation}

In the gauge used above,
$u_{\alpha0}(-k_F)=u_{\alpha0}(k_F)^*$.
We define
$S(V,J)=\sum_\alpha|u_{\alpha0}(k_F)|^4$.
We therefore have $\Gamma_{\mathrm D}=2US(V,J)$.
Including the equivalent process with the two incoming
branches interchanged cancels the factor $1/2$ in
Eq.~\eqref{eq:app-residual-quartic}. The continuum contribution is
\begin{equation}
H_{\mathrm D}
=a\Gamma_{\mathrm D}\int_0^\ell dx\,
:R^\dagger L^\dagger LR:_0
=\Gamma_{\mathrm D}\int_0^\ell dx\,:\rho_R\rho_L:_0.
\label{eq:app-direct-scattering}
\end{equation}

\paragraph{Exchange interbranch scattering.}

For the same incoming momenta near $k=k_F$ and $p=-k_F$, a transfer $q\simeq-2k_F$
exchanges the outgoing branches. Its vertex is
\begin{align}
\Gamma_{\mathrm X}
&=\Gamma(k_F,-k_F;-2k_F)
\nonumber\\
&=2U\cos(2k_F)\sum_\alpha
u_{\alpha0}^{*}(-k_F)u_{\alpha0}^{*}(k_F)
u_{\alpha0}(-k_F)u_{\alpha0}(k_F)=-2US(V,J).
\end{align}

Fermionic anticommutation gives
$ :L^\dagger R^\dagger LR:_0=-:\rho_R\rho_L:_0$,
so the exchange contribution is
\begin{equation}
H_{\mathrm X}
=-\Gamma_{\mathrm X}\int_0^\ell dx\,:\rho_R\rho_L:_0.
\label{eq:app-exchange-scattering}
\end{equation}

Thus, direct and exchange scattering contribute to the same
continuum operator, and $S(V,J)$ can be evaluated using Eq.~\eqref{eq:app-u0-Fermi}:
\begin{equation}
H_2
=H_{\mathrm D}+H_{\mathrm X}
=g_2\int_0^\ell dx\,:\rho_R\rho_L:_0, \qquad
g_2
=\Gamma_{\mathrm D}-\Gamma_{\mathrm X}
=4U\frac{V^4+8J^4}{(V^2+4J^2)^2}
\label{eq:app-g2-parameters}
\end{equation}

Using
$\rho_R=(-\partial_x\phi+\partial_x\theta)/(2\pi)$ and
$\rho_L=(-\partial_x\phi-\partial_x\theta)/(2\pi)$,
the bosonized interaction becomes
\begin{equation}
H_2=\frac{g_2}{4\pi^2}\int_0^\ell dx\,
\left[(\partial_x\phi)^2-(\partial_x\theta)^2\right].
\label{eq:app-g2-bosonized}
\end{equation}

\paragraph{Umklapp scattering.}

Specifically, near half filling, we consider the process $LL\rightarrow RR$, in which two
particles near the left Fermi point are scattered to the right
Fermi point. In the vertex $\Gamma(k,p;q)$, this corresponds to
\begin{equation}
k=-k_F+a\kappa_1,\qquad
p=-k_F+a\kappa_2,\qquad
q=2k_F+aQ,
\label{eq:app-umklapp-momenta}
\end{equation}

where $\kappa_1$, $\kappa_2$, and $Q$ are small physical momenta. In direct and exchange
scattering, these small momenta modify $g_2$ at $O(aQ)^2$, whereas the Umklapp 
process has an $O(aQ)$ contribution.
We set $k_F=\pi/2$.
The corresponding scattering is therefore

\begin{equation}
\begin{aligned}
\Gamma(-k_F+a\kappa_1,-k_F+a\kappa_2;2k_F+aQ)
&\simeq
2U\cos(2k_F+aQ)
\sum_\alpha
\left[u_{\alpha0}^{*}(k_F)u_{\alpha0}(-k_F)\right]^2\\
&=-2U\cos(aQ)\,T(V,J).
\end{aligned}
\label{eq:app-umklapp-vertex}
\end{equation}

Similarly, $T(V,J)$ can be evaluated using Eq.~\eqref{eq:app-u0-Fermi}:

\begin{equation}
T(V,J)
=\sum_\alpha
\left[u_{\alpha0}^{*}(k_F)u_{\alpha0}(-k_F)\right]^2=\frac{V^4-8J^4}{(V^2+4J^2)^2}.
\end{equation}

Substituting the Umklapp vertex into the four-fermion
interaction and using $\ell=La$ in the bulk normalization gives
\begin{equation}
\begin{aligned}
H_{umk}
=-\frac{UaT(V,J)}{\ell}
\sum_{\kappa_1,\kappa_2,Q}\cos(aQ)\,
:R^\dagger_{\kappa_1+Q}
R^\dagger_{\kappa_2-Q}
L_{\kappa_2}L_{\kappa_1}:_0
+\mathrm{H.c.}
\end{aligned}
\label{eq:app-umklapp-momentum-H}
\end{equation}

We use $\cos(aQ)=(e^{iaQ}+e^{-iaQ})/2$ and the Fourier convention
\begin{equation}
\begin{aligned}
\int dx\,
:R^\dagger(x)R^\dagger(x+\delta)
L(x+\delta)L(x):_0 &=
\frac{1}{\ell}
\sum_{r_1,r_2,p_1,p_2}
\delta_{r_1+r_2,p_1+p_2}\,
e^{i(p_2-r_2)\delta}
:R^\dagger_{r_1}R^\dagger_{r_2}L_{p_2}L_{p_1}:_0\\
&=
\frac{1}{\ell}
\sum_{\kappa_1,\kappa_2,Q}
e^{iQ\delta}
:R^\dagger_{\kappa_1+Q}
R^\dagger_{\kappa_2-Q}
L_{\kappa_2}L_{\kappa_1}:_0,
\end{aligned}
\label{eq:app-umklapp-fourier-identity}
\end{equation}

where the last equality follows from
$r_1=\kappa_1+Q$, $r_2=\kappa_2-Q$,
$p_1=\kappa_1$, and $p_2=\kappa_2$.
This yields
\begin{equation}
\begin{aligned}
H_{umk}
=-\frac{UaT(V,J)}{2}\int dx\,
\Bigl[
&:R^\dagger(x)R^\dagger(x+a)L(x+a)L(x):_0+ :R^\dagger(x)R^\dagger(x-a)L(x-a)L(x):_0
+\mathrm{H.c.}
\Bigr].\\
&=-UaT(V,J)\int dx\,
\Bigl[
:R^\dagger(x)L(x)
R^\dagger(x+a)L(x+a):_0
+\mathrm{H.c.}
\Bigr].
\end{aligned}
\end{equation}

The two spatial orientations give identical bulk integrals:
after shifting $x$ in the second term, exchanging both the
creation operators and the annihilation operators produces
two compensating minus signs.

With the phase convention $R^\dagger(x)L(x)\propto:e^{2i\phi(x)}:_0$, 
the separated product $:R^\dagger(x)L(x)R^\dagger(x+a)L(x+a):_0$ generates
 the vertex $:\exp\{2i\phi(x)+2i\phi(x+a)\}:_0$, up to a short-distance 
 matching factor. Introducing the bond midpoint $X=x+a/2$, we expand the 
 slowly varying field as 
 $\phi(X-a/2)+\phi(X+a/2)=2\phi(X)+(a^2/4)\partial_X^2\phi(X)+O(a^4)$. 
 Neglecting derivative corrections therefore gives $:e^{4i\phi(X)}:_0$, 
 which combines with its Hermitian conjugate to produce $2:\cos[4\phi(X)]:_0$.
  Choosing the matching scale $\alpha_0=a$ and absorbing the short-distance
   normalization into $g_u$, we obtain

\begin{equation}
H_u
=-\frac{g_u}{2\pi^2\alpha_0^2}
\int_0^\ell dx\,:\cos[4\phi(x)]:_0, \qquad
g_u
=UaT(V,J)
=Ua\,\frac{V^4-8J^4}{(V^2+4J^2)^2}.
\label{eq:app-umklapp-bosonized}
\end{equation}

Combining the three scattering processes, the retained
four-fermion contribution at the half-filled reference point is
\begin{equation}
H_{U}^{(4)}
\simeq{}
\frac{g_2}{4\pi^2}\int_0^\ell dx\,
\left[(\partial_x\phi)^2-(\partial_x\theta)^2\right]-\frac{g_u}{2\pi^2\alpha_0^2}
\int_0^\ell dx\,:\cos[4\phi(x)]:_0.
\label{eq:app-quartic-final}
\end{equation}

\subsection{Projection and bosonization of the pairing term}
\label{sec:app-pairing}

We next consider the uniform intraorbital pairing term from~\eqref{pair}.
After Fourier transformation, we have
\begin{equation}
H_\Delta
=\sum_{q,\alpha}
\left[
-i\Delta\sin q\,
c^\dagger_{q,\alpha}c^\dagger_{-q,\alpha}
+\mathrm{H.c.}
\right].
\label{eq:app-pairing-momentum}
\end{equation}

Using the orbital-to-band transformation gives
\begin{equation}
\begin{aligned}
H_\Delta
&=\sum_{q,s,s'}
\left[
-i\Delta\sin q\,\mathcal F_{ss'}(q)
d^\dagger_{q,s}d^\dagger_{-q,s'}
+\mathrm{H.c.}
\right],\\
\mathcal F_{ss'}(q)
&=\sum_\alpha
u_{\alpha s}^{*}(q)u_{\alpha s'}^{*}(-q).
\end{aligned}
\label{eq:app-pairing-band-basis}
\end{equation}

For the middle-band eigenvector used above~\eqref{eq:app-u0},
$u_{\alpha0}(-q)=u_{\alpha0}(q)^*$, and therefore
\begin{equation}
\mathcal F_{00}(q)
=\sum_\alpha|u_{\alpha0}(q)|^2=\frac{2|f(q)|^2+V^2}{\mathcal E(q)^2}=1.
\label{eq:app-pairing-form-factor}
\end{equation}

Thus, the uniform pairing amplitude is not reduced by a
middle-band orbital form factor. 
To obtain the continuum limit, we expand around $k_F=\pi/2$.
For $q=q_F+ap$, the pair-creation operator becomes
$d_q^\dagger d_{-q}^\dagger
\simeq R_p^\dagger L_{-p}^\dagger$.
The contribution from $q=-q_F+ap$ is identical after
reordering the creation operators. Consequently,
\begin{equation}
H_\Delta^{\mathrm{proj}}
\simeq
-2i\Delta
\sum_p R_p^\dagger L_{-p}^\dagger+\mathrm{H.c.}=
-2i\Delta
\int_0^\ell dx\,R^\dagger(x)L^\dagger(x)
+\mathrm{H.c.},
\label{eq:app-pairing-chiral}
\end{equation}

Under bosonization, the density-field factors cancel in
$R^\dagger L^\dagger$, leaving a vertex proportional to
$e^{-2i\theta}$. The pairing term then takes the form
\begin{equation}
H_\Delta^{\mathrm{proj}}
=-\frac{2\Delta}{\pi\alpha_0}
\int_0^\ell dx\,:\cos[2\theta(x)]:_0.
\label{eq:app-pairing-bosonized}
\end{equation}

Here $\alpha_0$ in $H_\Delta^{\mathrm{proj}}$ and $H_{u}$ above is the ultraviolet length cutoff of the
bosonized theory. For the lattice-scale matching used in our
numerical estimates, we choose $\alpha_0=a=1$ unit cell. 
A different cutoff convention requires a corresponding
rematching of these couplings and should not be implemented
by changing $\alpha_0$ alone.

\subsection{Low-energy Hamiltonian}
\label{sec:app-low-energy-H}

Combining the linearized kinetic energy~\eqref{noninteract}, the bilinear
corrections~\eqref{eq:app-bilinear-final}, the four-fermion interaction~\eqref{eq:app-quartic-final}, and the
projected pairing term~\eqref{eq:app-pairing-bosonized}, we obtain
\begin{equation}
\begin{aligned}
H_{\mathrm{eff}}
={}&\frac{v_F}{2\pi}\int_0^\ell dx\,
\left[(\partial_x\phi)^2+(\partial_x\theta)^2\right]+\frac{g_4}{4\pi^2}\int_0^\ell dx\,
\left[(\partial_x\phi)^2+(\partial_x\theta)^2\right]+\frac{g_2}{4\pi^2}\int_0^\ell dx\,
\left[(\partial_x\phi)^2-(\partial_x\theta)^2\right]\\
&+\frac{\mu_{\mathrm{eff}}}{\pi}\int_0^\ell dx\,\partial_x\phi-\frac{g_u}{2\pi^2\alpha_0^2}
\int_0^\ell dx\,:\cos[4\phi(x)]:_0-\frac{2\Delta}{\pi\alpha_0}
\int_0^\ell dx\,:\cos[2\theta(x)]:_0.
\end{aligned}
\label{eq:app-effective-Hamiltonian-expanded}
\end{equation}

The first three terms can be combined into a Gaussian
Tomonaga--Luttinger Hamiltonian by introducing the effective
Luttinger parameter $K$ and velocity $v$:
\begin{equation}
K=\sqrt{
\frac{2\pi v_F+g_4-g_2}
     {2\pi v_F+g_4+g_2}
} \qquad
v
=\sqrt{
\left(v_F+\frac{g_4}{2\pi}\right)^2
-\left(\frac{g_2}{2\pi}\right)^2
}.
\label{eq:app-K-v}
\end{equation}

The final low-energy Hamiltonian is
\begin{equation}
H_{\mathrm{eff}}
={}\frac{v}{2\pi}\int_0^\ell dx\,
\left[
K(\partial_x\theta)^2
+\frac{1}{K}(\partial_x\phi)^2
\right]
+\frac{\mu_{\mathrm{eff}}}{\pi}
\int_0^\ell dx\,\partial_x\phi
-\frac{g_u}{2\pi^2\alpha_0^2}
\int_0^\ell dx\,:\cos[4\phi(x)]:_0
-\frac{2\Delta}{\pi\alpha_0}
\int_0^\ell dx\,:\cos[2\theta(x)]:_0.
\label{eq:app-effective-Hamiltonian}
\end{equation}

The parameters are given by the following expressions: $g_4$ in Eqs.~\eqref{eq:app-g4-parameters} and~\eqref{eq:app-bg-selfenergy-derivative},
$g_2$ in Eq.~\eqref{eq:app-g2-parameters}, $\mu_{\mathrm{eff}}$ in Eqs.~\eqref{eq:app-mu-parameters}, \eqref{eq:app-bg-selfenergy-F}, and~\eqref{eq:app-bilinear-coefficients},
and $g_u$ in Eq.~\eqref{eq:app-umklapp-bosonized}.

The effective chemical potential provides a possible mechanism
for incommensurate density correlations. Completing the square
in the Gaussian density sector and introducing
$\widetilde\phi(x)=\phi(x)+K\mu_{\mathrm{eff}}x/v$
removes the term linear in $\partial_x\phi$, up to a constant,
while transforming the Umklapp potential into
$\cos[4\widetilde\phi(x)-Q_{\mathrm{mis}}x]$, where
$Q_{\mathrm{mis}}=4K\mu_{\mathrm{eff}}/v$.
\begin{equation}
H_\phi
=\int_0^\ell dx\left[
\frac{v}{2\pi K}(\partial_x\widetilde\phi)^2
-\lambda_u\cos(4\widetilde\phi-Q_{\mathrm{mis}}x)
\right]+\mathrm{const}
\end{equation}

In the soliton description, the gradient favors $\partial_x\widetilde\phi=0$, while locking by the cosine requires its argument
$4\widetilde\phi-Q_{\mathrm{mis}}x$ to remain at a minimum
of the cosine potential, $ie$. $\partial_x\widetilde\phi=K\mu_{\mathrm{eff}}/v$.
Remaining on the same branch of minima of the cosine potential
incurs a gradient-energy cost, whereas a transition to an adjacent
branch introduces a soliton. Therefore a finite soliton density can move the
density modulation away from commensurability. Provided that
the pairing perturbation does not gap the resulting density
mode, this mechanism supports an incommensurate density-wave
regime.

\subsection{Renormalization-group analysis}
\label{sec:app-RG}

We derive the bulk renormalization-group (RG) equations to
second order in the nonlinear couplings. 
The derivation assumes weak couplings and retains
the leading pairing and Umklapp operators together with
the Gaussian stiffnesses and the incommensurate mismatch. After shifting the density field to eliminate the initial
chemical-potential term by
$Q_{\mathrm{mis}}=4K\mu_{\mathrm{eff}}/v$, the Hamiltonian reads

\begin{equation}
H_{\mathrm{eff}}
={}\frac{v}{2\pi}\int dx\,
\left[
K(\partial_x\theta)^2+
\frac{1}{K}(\partial_x\widetilde\phi)^2
\right]
-\frac{g_u}{2\pi^2\alpha^2}
\int dx\,:\cos(4\widetilde\phi-Q_{\mathrm{mis}}x):
-\frac{2\Delta}{\pi\alpha}
\int dx\,:\cos(2\theta):,
\label{eq:app-RG-start}
\end{equation}

We introduce the dimensionless couplings
\begin{equation}
\delta=\frac{4\alpha\Delta}{v},
\qquad
y=\frac{g_u}{\pi v},
\qquad
\zeta=\alpha Q_{\mathrm{mis}},
\label{eq:app-RG-couplings}
\end{equation}
and the RG parameter $l=\ln(\alpha/\alpha_0)$.
Using the Euclidean coordinate $z=v\tau$, the perturbation
to the Gaussian action becomes
\begin{equation}
\begin{aligned}
S&=S_0+S_{\mathrm{pert}},\\
S_0
&=\frac{1}{2\pi K}\int dx\,dz\,
\left[
(\partial_x\widetilde\phi)^2
+(\partial_z\widetilde\phi)^2
\right]\\
S_{\mathrm{pert}}
&=-\frac{1}{2\pi\alpha^2}\int dx\,dz\,
\left[
y\cos(4\widetilde\phi-Q_{\mathrm{mis}}x)
+\delta\cos(2\theta)
\right]
\end{aligned}
\label{eq:app-RG-action}
\end{equation}

Following the standard renormalization procedure in Refs.~\cite{giamarchi2004} and~\cite{kardarStatisticalPhysicsFields2007},
we decompose the fields into fast modes ($>$), with
$\Lambda e^{-dl}<|\mathbf p|<\Lambda$, and slow modes ($<$),
with $|\mathbf p|<\Lambda e^{-dl}$. Integrating the Gaussian propagators over the momentum shell
$\Lambda e^{-dl}<|\mathbf p|<\Lambda$ gives the
equal-point fast-mode variances
$\langle\widetilde\phi_>^2(\mathbf r)\rangle_>=K\,dl/2$
and $\langle\theta_>^2(\mathbf r)\rangle_>=dl/(2K)$. After integrating out
the fast modes, the effective action for the slow modes to second order is
\begin{equation}
S_{\mathrm{eff}}(<)
={}S_0(<)
+\langle S_{\mathrm{pert}}\rangle_>-\frac12
\left[
\langle S_{\mathrm{pert}}^2\rangle_>
-\langle S_{\mathrm{pert}}\rangle_>^2
\right]
+O(S_{\mathrm{pert}}^3).
\label{eq:app-cumulant-expansion}
\end{equation}

For the first-order correction $\langle S_{\mathrm{pert}}\rangle_>$, the Umklapp vertex gives
\begin{equation}
\begin{aligned}
&\left\langle
\cos\!\left(
4\widetilde\phi_<+4\widetilde\phi_>
-Q_{\mathrm{mis}}x
\right)
\right\rangle_> =
\operatorname{Re}\left[
e^{i(4\widetilde\phi_<-Q_{\mathrm{mis}}x)}
\left\langle e^{4i\widetilde\phi_>}\right\rangle_>
\right]\\
&\quad=
e^{-8\langle\widetilde\phi_>^2\rangle_>}
\cos(4\widetilde\phi_<-Q_{\mathrm{mis}}x)=
e^{-4K\,dl}
\cos(4\widetilde\phi_<-Q_{\mathrm{mis}}x).
\end{aligned}
\label{eq:app-umklapp-fast-average}
\end{equation}

Similarly, for the pairing vertex,
\begin{equation}
\begin{aligned}
&\left\langle
\cos(2\theta_<+2\theta_>)
\right\rangle_>
=\operatorname{Re}\left[
e^{2i\theta_<}
\left\langle e^{2i\theta_>}\right\rangle_>
\right]\\
&=e^{-2\langle\theta_>^2\rangle_>}
\cos(2\theta_<)=e^{-dl/K}\cos(2\theta_<).
\end{aligned}
\label{eq:app-pairing-fast-average}
\end{equation}

After integrating out the fast modes, the momentum cutoff is
reduced to $\Lambda e^{dl}$. We restore the original
cutoff by rescaling the coordinates and momenta as $\mathbf r=b\mathbf r'$, and the fields are $\widetilde\phi'(\mathbf r')=\widetilde\phi_<(b\mathbf r')$
and $\theta'(\mathbf r')=\theta_<(b\mathbf r')$.
The coordinate and gradient rescaling factors in $S_0$ cancel, giving
\begin{equation}
\langle S_{\mathrm{pert}}\rangle_>
=-\frac{1}{2\pi\alpha^2}\int d^2\mathbf r'\,
\Bigl[
y\,b^{2-4K}
\cos\!\left(4\widetilde\phi'-bQ_{\mathrm{mis}}x'\right)
+\delta\,b^{2-1/K}\cos(2\theta')
\Bigr].
\end{equation}

Comparing with the original action gives
\begin{equation}
y(l+dl)=y(l)e^{[2-4K(l)]dl},
\qquad
\delta(l+dl)=\delta(l)e^{[2-1/K(l)]dl}.
\end{equation}

Expanding to first order in $dl$, we obtain the RG flow of $y$ and $\delta$
\begin{equation}
\frac{dy}{dl}=(2-4K)y,
\qquad
\frac{d\delta}{dl}=\left(2-\frac{1}{K}\right)\delta.
\label{eq:app-RG-linear}
\end{equation}

For $y$ and $\delta$ within second-order RG, the second cumulant generates Gaussian
corrections and higher-order vertices, but does not reproduce
either original cosine. The resulting renormalization of $K$
is retained through $K(l)$, whereas direct nonlinear corrections
to the amplitude flows begin at
$O(y^3,y\delta^2)$ and $O(\delta^3,\delta y^2)$, respectively,
and lie beyond the present truncation. Therefore, we retain the flow equations for these two couplings only 
to first order in their amplitudes.

The second-order correction is
\begin{equation}
\Delta S^{(2)}
=-\frac12
\left[
\langle S_{\mathrm{pert}}^2\rangle_>
-\langle S_{\mathrm{pert}}\rangle_>^2
\right]
=\Delta S_{uu}
+\Delta S_{\Delta\Delta}
+\Delta S_{u\Delta}.
\label{eq:app-second-cumulant}
\end{equation}

For the Gaussian fast modes, the connected exponential
averages are defined as $\langle O_1O_2\rangle_{c}=\langle O_1O_2\rangle -\langle O_1\rangle\langle O_2\rangle$.
Using the Gaussian identity 
$\left\langle e^{i(aX+bY)}\right\rangle=\exp\!\left[-\frac12\left(
a^2\langle X^2\rangle
+b^2\langle Y^2\rangle
+2ab\langle XY\rangle
\right)
\right]$, we have

\begin{align}
&\left\langle
e^{4i\sigma\widetilde\phi_{>,1}}
e^{4i\tau\widetilde\phi_{>,2}}
\right\rangle_{>,c}
=e^{-8K\,dl}
\left[
e^{-16\sigma\tau G_{\phi\phi}^{>}(\mathbf s)}-1
\right],
\label{eq:app-connected-phi-vertices}\\
&\left\langle
e^{2i\sigma\theta_{>,1}}
e^{2i\tau\theta_{>,2}}
\right\rangle_{>,c}
=e^{-2dl/K}
\left[
e^{-4\sigma\tau G_{\theta\theta}^{>}(\mathbf s)}-1
\right],
\label{eq:app-connected-theta-vertices}\\
&\left\langle
e^{4i\sigma\widetilde\phi_{>,1}}
e^{2i\tau\theta_{>,2}}
\right\rangle_{>,c}
=e^{-(4K+1/K)dl}
\left[
e^{-8\sigma\tau G_{\phi\theta}^{>}(\mathbf s)}-1
\right].
\label{eq:app-connected-mixed-vertices}
\end{align}

where $\sigma,\tau=\pm1$. The fast-mode covariances are defined by
\begin{equation}
G_{\phi\phi}^{>}(\mathbf s)
\equiv
\left\langle
\widetilde\phi_>(\mathbf r_1)
\widetilde\phi_>(\mathbf r_2)
\right\rangle_>,  \quad
G_{\theta\theta}^{>}(\mathbf s)
\equiv
\left\langle
\theta_>(\mathbf r_1)\theta_>(\mathbf r_2)
\right\rangle_>, \quad
G_{\phi\theta}^{>}(\mathbf s)
\equiv
\left\langle
\widetilde\phi_>(\mathbf r_1)\theta_>(\mathbf r_2)
\right\rangle_>,
\label{eq:app-fast-covariances}
\end{equation}
where $s=r_1-r_2$ with Euclidean time ordering understood.
For the circular momentum shell
$\Lambda e^{-dl}<|\mathbf p|<\Lambda$, the diagonal
covariances are
\begin{equation}
\begin{aligned}
G_{\phi\phi}^{>}(\mathbf s)
&=\int_{\mathrm{shell}}\frac{d^2\mathbf p}{(2\pi)^2}
\frac{\pi K}{|\mathbf p|^2}e^{i\mathbf p\cdot\mathbf s}
=\frac{K}{2}\int_{\Lambda e^{-dl}}^\Lambda
\frac{dp}{p}\,J_0(p|\mathbf s|),\\
G_{\theta\theta}^{>}(\mathbf s)
&=\int_{\mathrm{shell}}\frac{d^2\mathbf p}{(2\pi)^2}
\frac{\pi}{K|\mathbf p|^2}e^{i\mathbf p\cdot\mathbf s}
=\frac{1}{2K}\int_{\Lambda e^{-dl}}^\Lambda
\frac{dp}{p}\,J_0(p|\mathbf s|).
\end{aligned}
\end{equation}

Here $J_0$ denotes the Bessel function of the first kind
of order zero. In particular,
$G_{\phi\phi}^{>}(0)=K\,dl/2$ and
$G_{\theta\theta}^{>}(0)=dl/(2K)$.
We substitute these expressions into the second cumulant. For example, writing
$A_1=\frac12\sum_{\sigma=\pm1}
e^{i\sigma[4\widetilde\phi_{<,1}-Q_{\mathrm{mis}}x_1]}
e^{4i\sigma\widetilde\phi_{>,1}}$
and
$A_2=\frac12\sum_{\tau=\pm1}
e^{i\tau[4\widetilde\phi_{<,2}-Q_{\mathrm{mis}}x_2]}
e^{4i\tau\widetilde\phi_{>,2}}$,
the connected average becomes
\begin{equation}
\langle A_1A_2\rangle_{>,c}
={}\frac14\sum_{\sigma,\tau=\pm1}
e^{-iQ_{\mathrm{mis}}(\sigma x_1+\tau x_2)}
e^{4i\sigma\widetilde\phi_{<,1}
+4i\tau\widetilde\phi_{<,2}} \times\Bigl[
\left\langle
e^{4i\sigma\widetilde\phi_{>,1}}
e^{4i\tau\widetilde\phi_{>,2}}
\right\rangle_>
-
\left\langle e^{4i\sigma\widetilde\phi_{>,1}}\right\rangle_>
\left\langle e^{4i\tau\widetilde\phi_{>,2}}\right\rangle_>
\Bigr].
\end{equation}

\begin{align}
\Delta S_{uu}
={}&-\frac{y^2e^{-8K\,dl}}{32\pi^2\alpha^4}
\sum_{\sigma,\tau=\pm1}
\int d^2\mathbf r_1\,d^2\mathbf r_2\,
e^{-iQ_{\mathrm{mis}}(\sigma x_1+\tau x_2)}
e^{4i\sigma\widetilde\phi_{<,1}
  +4i\tau\widetilde\phi_{<,2}}
\times
\left[
e^{-16\sigma\tau G_{\phi\phi}^{>}(\mathbf s)}-1
\right],
\label{eq:app-uu-exponential}\\
\Delta S_{\Delta\Delta}
={}&-\frac{\delta^2e^{-2dl/K}}{32\pi^2\alpha^4}
\sum_{\sigma,\tau=\pm1}
\int d^2\mathbf r_1\,d^2\mathbf r_2\,
e^{2i\sigma\theta_{<,1}+2i\tau\theta_{<,2}}
\times
\left[
e^{-4\sigma\tau G_{\theta\theta}^{>}(\mathbf s)}-1
\right],
\label{eq:app-DD-exponential}\\
\Delta S_{u\Delta}
={}&-\frac{y\delta\,e^{-(4K+1/K)dl}}{16\pi^2\alpha^4}
\sum_{\sigma,\tau=\pm1}
\int d^2\mathbf r_1\,d^2\mathbf r_2\,
e^{-i\sigma Q_{\mathrm{mis}}x_1}
e^{4i\sigma\widetilde\phi_{<,1}
  +2i\tau\theta_{<,2}}
\times
\left[
e^{-8\sigma\tau G_{\phi\theta}^{>}(\mathbf s)}-1
\right].
\label{eq:app-uD-exponential}
\end{align}

The integral can be separated into $R=(r_1+r_2)/2$ and $s=r_1-r_2$. We now extract
the contributions proportional to $(\partial \phi)^2$. The key term is $e^{4i\sigma\widetilde\phi_{<,1}
  +2i\tau\theta_{<,2}}$. In $S_{u,\Delta}$,  this term cannot cancel when choosing
$\sigma ,\tau = \pm 1$, leaving the term $\cos(4\phi+2\theta)(\nabla\phi) \neq (\nabla\phi)^2.$. Therefore,
this term cannot renormalize $K$ at second order. For the other two terms,
integrating over the circular short-distance shell
$\alpha<|\mathbf s|<\alpha e^{dl}$ gives the Gaussian
corrections
\begin{align}
\Delta S_{uu}
={}&\frac{y^2dl}{\pi}\int d^2\mathbf R\,
\Bigl[
\bigl(J_0(\zeta)-J_2(\zeta)\bigr)
(\partial_x\widetilde\phi_<)^2
+\bigl(J_0(\zeta)+J_2(\zeta)\bigr)
(\partial_z\widetilde\phi_<)^2
\Bigr]-\frac{y^2}{\pi\alpha}
J_1(\zeta)\,dl
\int dx\,dz\,\partial_x\widetilde\phi_<,
\label{eq:app-uu-integrated}\\
\Delta S_{\Delta\Delta}^{\mathrm{grad}}
={}&\frac{\delta^2dl}{4\pi}\int d^2\mathbf R\,
\left[
(\partial_x\theta_<)^2
+(\partial_z\theta_<)^2
\right],
\label{eq:app-DD-integrated}
\end{align}

where $\zeta=Q_{\mathrm{mis}}\alpha$ and $J_n$ denotes
the Bessel function of the first kind of order $n$. The pairing correction is first combined with the Gaussian
action in the dual representation:
\begin{equation}
\begin{aligned}
S_0[\theta]+dS_{\Delta\Delta}^{\mathrm{grad}}
&=\frac{1}{2\pi}\int d^2\mathbf R\,
\left(K+\frac{\delta^2}{2}dl\right)
(\nabla\theta_<)^2.
\end{aligned}
\end{equation}

Thus, pairing produces $dK_\Delta=\delta^2dl/2$.
Expressed in the $\widetilde\phi$ representation~\eqref{eq:app-RG-action}, the
corresponding inverse stiffness is
\begin{equation}
\frac{1}{K+dK_\Delta}
=\frac1K-\frac{\delta^2}{2K^2}dl+O(dl^2).
\end{equation}

Combining this correction with the Umklapp contribution
therefore gives
\begin{equation}
\begin{aligned}
S_{\mathrm{G,eff}}
=\frac{1}{2\pi}\int dx\,dz\,
\Biggl\{
&\left[
\frac1K
+2y^2\bigl(J_0(\zeta)-J_2(\zeta)\bigr)dl
-\frac{\delta^2}{2K^2}dl
\right]
(\partial_x\widetilde\phi_<)^2\\
+&\left[
\frac1K
+2y^2\bigl(J_0(\zeta)+J_2(\zeta)\bigr)dl
-\frac{\delta^2}{2K^2}dl
\right]
(\partial_z\widetilde\phi_<)^2
\Biggr\},
\end{aligned}
\label{eq:app-renormalized-Gaussian-action}
\end{equation}

To identify the running parameters, write the renormalized
Gaussian action in the original imaginary-time coordinate:
\begin{equation}
\begin{aligned}
S_{\mathrm{G,eff}}
={}&\frac{1}{2\pi}\int dx\,d\tau\,
\left[
\frac{v+dv}{K+dK}(\partial_x\widetilde\phi_<)^2
+\frac{1}{(v+dv)(K+dK)}
(\partial_\tau\widetilde\phi_<)^2
\right]\\
\simeq{}&\frac{1}{2\pi}\int dx\,dz\,
\Biggl[
\left(
\frac1K-\frac{dK}{K^2}+\frac{dv}{Kv}
\right)(\partial_x\widetilde\phi_<)^2
+\left(
\frac1K-\frac{dK}{K^2}-\frac{dv}{Kv}
\right)(\partial_z\widetilde\phi_<)^2
\Biggr]
\end{aligned}
\label{eq:app-Gaussian-parameter-expansion}
\end{equation}

Comparing these two terms yields
\begin{equation}
\begin{aligned}
-\frac{dK}{K^2}+\frac{dv}{Kv}
&=\left[
2y^2\bigl(J_0(\zeta)-J_2(\zeta)\bigr)
-\frac{\delta^2}{2K^2}
\right]dl,\\
-\frac{dK}{K^2}-\frac{dv}{Kv}
&=\left[
2y^2\bigl(J_0(\zeta)+J_2(\zeta)\bigr)
-\frac{\delta^2}{2K^2}
\right]dl.
\end{aligned}
\end{equation}

Therefore, the resulting second-order RG equations are
\begin{equation}
\frac{dK}{dl}
=\frac{\delta^2}{2}-2K^2y^2J_0(\zeta),
\qquad
\frac{dv}{dl}
=-2vKy^2J_2(\zeta),
\qquad
\label{eq:app-K-v-RG}
\end{equation}

From ~\eqref{eq:app-uu-integrated}
at nonzero $\zeta$, the term linear in the slow-field
gradient generated by the Umklapp cumulant is
\begin{equation}
dS_{\mathrm{lin}}
=-\frac{y^2J_1(\zeta)}{\pi\alpha}\,dl
\int dx\,dz\,\partial_x\widetilde\phi.
\label{eq:app-RG-generated-gradient}
\end{equation}

This term is a correction to the chemical-potential shift
\begin{equation}
\begin{aligned}
\widetilde\phi_{\mathrm{new}}
&=\widetilde\phi_<
-\frac{Ky^2}{\alpha}J_1(\zeta)\,dl\,x,\\
dQ_{\mathrm{mis}}
&=-\frac{4Ky^2}{\alpha}J_1(\zeta)\,dl.
\end{aligned}
\label{eq:app-field-shift-misfit}
\end{equation}

Including the engineering scaling of
$\zeta=\alpha Q_{\mathrm{mis}}$ yields
\begin{equation}
\frac{d\zeta}{dl}
=\zeta-4Ky^2J_1(\zeta).
\label{eq:app-RG-mismatch}
\end{equation}

Collecting the contributions from Eqs.~\eqref{eq:app-RG-linear}, \eqref{eq:app-K-v-RG}, and~\eqref{eq:app-RG-mismatch}, 
the second-order RG flow equations are
\begin{equation}
\boxed{
\begin{aligned}
\frac{dK}{dl}
&=\frac{\delta^2}{2}
-2K^2y^2J_0(\zeta),\\
\frac{d\delta}{dl}
&=\left(2-\frac1K\right)\delta,\\
\frac{dy}{dl}
&=(2-4K)y,\\
\frac{d\zeta}{dl}
&=\zeta-4Ky^2J_1(\zeta),\\
\frac{dv}{dl}
&=-2vKy^2J_2(\zeta).
\end{aligned}
}
\label{eq:app-RG-full}
\end{equation}

The dimensionless Umklapp, pairing, and misfit parameters
are defined as
\begin{equation}
y=\frac{g_u}{\pi v},
\qquad
\delta=\frac{4\alpha\Delta}{v},
\qquad
\zeta=Q_{\mathrm{mis}}\alpha.
\end{equation}

\subsection{IC-C phase transition point}
\label{app:reduced-RG-stability}

To isolate the competition between commensurate locking and
incommensurability, we set the pairing coupling $\delta=0$.
Pairing is irrelevant in the ICDW and CDW regimes, motivated
by the observed exponential decay of connected pairing correlations
and the independently established gapless character of the ICDW phase.
The remaining flows are
\begin{equation}
\begin{aligned}
\frac{dK}{dl}
&=-2K^2y^2J_0(\zeta),\\
\frac{dy}{dl}
&=(2-4K)y,\\
\frac{d\zeta}{dl}
&=\zeta-4Ky^2J_1(\zeta),
\end{aligned}
\label{eq:app-reduced-IC-RG}
\end{equation}

The RG equations admit a Gaussian fixed line and additional
nonzero fixed points. We select the marginal Gaussian point
at $K_*=1/2$. The two fixed points considered here follow from Eq.~\eqref{eq:app-reduced-IC-RG}:
\begin{equation}
\begin{aligned}
\mathcal F_1:\quad
(K_*,y_*,\zeta_*)&=\left(\frac12,0,0\right),\\
\mathcal F_2:\quad
(K_*,y_*,\zeta_*)&=
\left(\frac12,-1.52,-2.40\right).
\end{aligned}
\label{eq:app-fixedpoints}
\end{equation}

To discuss the stability of these two fixed points, we
consider small deviations from the fixed point
$\mathbf g_*=(K_*,Y_*,\zeta_*)^{\mathsf T}$.
The RG equations are
\begin{equation}
\frac{d\mathbf g}{dl}
=\boldsymbol{\beta}(\mathbf g),
\qquad
\mathbf g=
\begin{pmatrix}
K\\Y\\\zeta
\end{pmatrix},
\qquad
\boldsymbol{\beta}(\mathbf g)=
\begin{pmatrix}
-2K^2Y^2J_0(\zeta)\\
(2-4K)Y\\
\zeta-4KY^2J_1(\zeta)
\end{pmatrix}.
\label{eq:app-vector-RG}
\end{equation}

A fixed point satisfies $\boldsymbol{\beta}(\mathbf g_*)=0$.
Its stability is determined by whether a small displacement
from $\mathbf g_*$ decreases or increases under the RG flow. We write
$\mathbf g(l)=\mathbf g_*+\boldsymbol{\epsilon}(l)$, where
$\boldsymbol{\epsilon}(l)=c\,e^{\lambda l}\mathbf e$.
Since $l\to+\infty$ corresponds to the long-distance limit,
$\operatorname{Re}\lambda>0$ implies a growing perturbation
and hence an unstable fixed point, whereas
$\operatorname{Re}\lambda<0$ implies a decaying perturbation
and a stable fixed point. 
We expand each beta function as
\begin{equation}
\beta_i(\mathbf g_*+\boldsymbol{\epsilon})
={}\sum_j(M_*)_{ij}\epsilon_j+O(\epsilon^2),
\qquad
(M_*)_{ij}
=
\left.
\frac{\partial\beta_i}{\partial g_j}
\right|_{\mathbf g_*}.
\label{eq:app-beta-linearization}
\end{equation}

Consequently, the linearized flow is
\begin{equation}
\frac{d\boldsymbol{\epsilon}}{dl}
=M_*\boldsymbol{\epsilon}+O(\epsilon^2).
\label{eq:app-linearized-flow}
\end{equation}

The stability matrix  reads
\begin{equation}
M_*=
\begin{pmatrix}
-4K_*Y_*^2J_0(\zeta_*)
&
-4K_*^2Y_*J_0(\zeta_*)
&
2K_*^2Y_*^2J_1(\zeta_*)
\\
-4Y_*&2-4K_*&0
\\
-4Y_*^2J_1(\zeta_*)
&
-8K_*Y_*J_1(\zeta_*)
&
1-4K_*Y_*^2J_1'(\zeta_*)
\end{pmatrix}.
\label{eq:app-stability-matrix-general}
\end{equation}

For the fixed points $\mathcal F_1$ and $\mathcal F_2$, the stability matrices are
\begin{equation}
M_1=
\begin{pmatrix}
0&0&0\\
0&0&0\\
0&0&1
\end{pmatrix},
\qquad
M_2\simeq
\begin{pmatrix}
0&0&0.60\\
-6.09&0&0\\
-4.81&-3.16&2
\end{pmatrix}.
\end{equation}

Their eigenvalues are
\begin{equation}
\operatorname{spec}(M_1)=\{0,0,1\},
\qquad
\operatorname{spec}(M_2)
\simeq
\{2.60,\,-0.30+2.09i,\,-0.30-2.09i\}.
\end{equation}

The linearized solutions are
\begin{equation}
\begin{aligned}
F_1:\quad
\begin{pmatrix}
\epsilon_Y(l)\\
\epsilon_\zeta(l)
\end{pmatrix}
&=
\begin{pmatrix}
\epsilon_Y(0)\\
e^l\epsilon_\zeta(0)
\end{pmatrix},
\\[1.5ex]
F_2:\quad
\begin{pmatrix}
\epsilon_1(l)\\
\epsilon_2(l)
\end{pmatrix}
&=
e^{-0.30l}
\begin{pmatrix}
\cos(2.09l)&-\sin(2.09l)\\
\sin(2.09l)&\cos(2.09l)
\end{pmatrix}
\begin{pmatrix}
\epsilon_1(0)\\
\epsilon_2(0)
\end{pmatrix}.
\end{aligned}
\label{eq:app-fixedpoint-perturbations}
\end{equation}

Here $\epsilon_1$ and $\epsilon_2$ are real coordinates
in the subspace of $F_2$, obtained from linear
combinations of $(\epsilon_K,\epsilon_Y,\epsilon_\zeta)$.
Their stable amplitude satisfies
\begin{equation}
\epsilon_1(l)^2+\epsilon_2(l)^2
=
e^{-0.60l}
\left[\epsilon_1(0)^2+\epsilon_2(0)^2\right].
\end{equation}

This analysis helps assess which fixed point is
accessible from our microscopic initial conditions.
Near $F_1$, the incommensurability obeys
$d\epsilon_\zeta/dl=\epsilon_\zeta$.
Consequently, a trajectory with nonzero initial
incommensurability cannot approach $F_1$, which requires
$\zeta(l)\to0$. By contrast, $F_2$ has two stable
directions. With $U=U_c$, this initial condition can allow
RG flow towards the fixed point.

We can then solve Eq.~\eqref{eq:app-reduced-IC-RG} with the fixed point $F_2$.
These coupled differential equations can be solved numerically using the 
fourth-order Runge–Kutta method.  The initial parameters are all functions
of $U$:

\begin{equation}
\begin{aligned}
K_0(U)
&=
\sqrt{
\frac{2\pi v_F+g_4(U)-g_2(U)}
     {2\pi v_F+g_4(U)+g_2(U)}
},\\
v_0(U)
&=
\sqrt{
\left[v_F+\frac{g_4(U)}{2\pi}\right]^2
-\left[\frac{g_2(U)}{2\pi}\right]^2
},\\
y_0(U)
&=\frac{g_u(U)}{\pi v_0(U)},\\
\zeta_0(U)
&=\frac{4\alpha_0K_0(U)\mu_{\mathrm{eff}}(U)}{v_0(U)}.
\end{aligned}
\label{eq:app-IC-RG-initial}
\end{equation}

For $J=1$, $V=0.6$, $t=0.03$, and $\alpha_0=a=1$, 
the resulting phase transition point and effective Luttinger parameter are
\begin{equation}
U_c\simeq0.27, \qquad K_{\mathrm{eff},c}\simeq0.47,
\label{eq:app-IC-RG-critical-estimate}
\end{equation}

Within the proposed low-energy description, the TSC phase is a
gapped superconducting regime in which pairing locks the dual
field and supports localized Majorana boundary modes. Upon
crossing the TSC--IC transition, the system enters a gapless
incommensurate regime, and the bulk gap protecting the Majorana
modes is lost. Motivated by the numerical observations, we
describe this ICDW regime by an effective theory in which the
pairing perturbation is irrelevant. We are unable to locate the TSC--IC transition
point in this theory because its determination might require
a higher-order RG approximation. The effective Luttinger
parameter decreases rapidly from the TSC--IC boundary, where fitting in the 
gapped regime fails, to approximately $1/2$ near the
numerically identified IC--C boundary, consistent with our prediction
of the effective Luttinger parameter. 

In this interpretation, incommensurability originates from
the interaction-induced effective chemical potential generated
by the multiorbital projection, provided that no exact
particle--hole symmetry enforces its cancellation. At the IC–C transition,
the system returns to half filling, meaning that Umklapp scattering is 
relevant and doping is irrelevant. The reported abrupt recovery of half filling 
differs from the conventional dilute-soliton C--IC scenario, in which
the soliton density continuously vanishes upon approaching
the commensurate phase. 

Within our truncated RG equations,
the dilute-soliton C--IC point with $(y_*,\zeta_*)=(0,0)$ cannot be
approached from nonzero incommensurability, where doping exists
just to the left of the transition, creating dilute solitons, whereas the Umklapp 
 coupling $y$ exists just to its right. The negative fixed point
in Eqs.~\eqref{eq:app-IC-RG-critical-estimate} and~\eqref{eq:app-fixedpoints}, denoted by $F_2$, instead
provides a competition between Umklapp and incommensurability
couplings. Both doping and Umklapp scattering remain 
strong throughout the parameter regime of interest.
 In the ICDW phase, the former dominates, resulting in 
 a finite deviation from half filling; in the CDW phase, 
 the latter dominates, locking the system at half filling.

\section{Mean-field analysis of the interacting quantum metric length and Majorana instability}
\label{app:interacting-QML}

In this section, we investigate how repulsive interactions modify
the geometric localization scale of Majorana boundary modes.
We first obtain a Hartree estimate of the interaction-renormalized
quantum metric length (QML), using the orbital occupations measured
in the interacting ground state. We then incorporate this scale
into the two-component Majorana wavefunction and derive its
consequences for the hybridization between opposite boundaries.

\subsection{Hartree correction to the quantum metric length}

In the Hartree approximation, we use the interaction to replace $V$ with $V_H(U)$. Then we substitute
$V_H(U)$ into the noninteracting QML expression to obtain the interacting QML. Consider the interaction
\begin{equation}
H_U
=
U\sum_{j=1}^{L-1}\sum_{\alpha=A,B,C}
\left(n_{j\alpha}-\frac12\right)
\left(n_{j+1,\alpha}-\frac12\right),
\label{eq:IQML-interaction}
\end{equation}

with open boundary conditions and zero bare chemical potential.
The orbital potentials are chosen as
\begin{equation}
H_V
=
V\sum_j(n_{jA}-n_{jC}),
\qquad V>0.
\label{eq:IQML-orbital-convention}
\end{equation}

Thus, the $A$ and $C$ orbitals have bare energies $+V$ and $-V$.
Writing
$\delta n_{j\alpha}=n_{j\alpha}-1/2$ and
$\overline{\delta n}_{j\alpha}
=\langle n_{j\alpha}\rangle-1/2$,
the Hartree decoupling is
\begin{equation}
H_U^{\mathrm H}
=
\sum_{j,\alpha}\Sigma_{j\alpha}^{\mathrm H}n_{j\alpha}
+E_{\mathrm H},
\end{equation}

where the open-chain Hartree self-energy is
\begin{equation}
\Sigma_{j\alpha}^{\mathrm H}
=U\Biggl[
\mathbf 1_{j>1}
\left(\langle n_{j-1,\alpha}\rangle-\frac12\right)
+\mathbf 1_{j<L}
\left(\langle n_{j+1,\alpha}\rangle-\frac12\right)
\Biggr].
\label{eq:IQML-local-Hartree}
\end{equation}

In the approximately homogeneous bulk, only the $A$ and $C$ orbitals contribute to $V$.
Because the QML is an even function of the relative Hartree shift of the \(B\) orbital,
its direct contribution starts at second order and can therefore be neglected in the first-order Hartree estimate of the QML.
The renormalized half-splitting of the $A$ orbital relative to the $C$ orbital is
therefore
\begin{equation}
V_{\mathrm H}
=
\frac{
[V+\Sigma_A^{\mathrm H}]
-[-V+\Sigma_C^{\mathrm H}]
}{2}
=
V-U(\overline n_C-\overline n_A)
\label{eq:IQML-VH}
\end{equation}

Defining $m\equiv\overline n_C-\overline n_A$, the corresponding
interaction-renormalized scale in the reduced Hartree Hamiltonian is obtained by substituting
$V_{\mathrm H}=V-Um$ into the noninteracting QML expression~\eqref{eq:IQML-bare}
\begin{equation}
\xi_{\mathrm{QML}}^{\mathrm H}(U,V)
=
\frac{
2aJ^2(V_{\mathrm H}^2+4J^2)
}{
V_{\mathrm H}(V_{\mathrm H}^2+8J^2)^{3/2}
},
\label{eq:IQML-substitution}
\end{equation}

To obtain the leading correction used in our numerical comparison,
we expand around the bare value of $V$:
\begin{equation}
\xi_{\mathrm{QML}}^{\mathrm{MF}}
\simeq
\xi_{\mathrm{QML}}^{(0)}
-Um\,\partial_V\xi_{\mathrm{QML}}^{(0)}.
\end{equation}

We obtain
\begin{equation}
\xi_{\mathrm{QML}}^{\mathrm{MF}}(U,V)
\simeq
\xi_{\mathrm{QML}}^{(0)}(V,J)
\left[
1+\frac{U(\overline n_C-\overline n_A)}{V}\mathcal A(V,J)
\right],
\label{eq:IQML-linear}
\end{equation}

where
\begin{equation}
\mathcal A(V,J)
=
\frac{2(V^4+4J^2V^2+16J^4)}
{(V^2+8J^2)(V^2+4J^2)}
\end{equation}

Thus, a positive orbital polarization $m$ produces a positive
leading correction to the QML. In the small-$V/J$ limit,
\begin{equation}
\xi_{\mathrm{QML}}^{(0)}
\simeq
\frac{aJ}{2\sqrt{2}V},
\qquad
\mathcal A(V,J)\longrightarrow1,
\end{equation}

and hence
\begin{equation}
\xi_{\mathrm{QML}}^{\mathrm{MF}}(U,V)
\simeq
\xi_{\mathrm{QML}}^{(0)}(V)
\left[
1+\frac{U}{V}
\left(\bar n_C-\bar n_A\right)
\right].
\label{eq:IQML-small-V}
\end{equation}

Therefore, within the Hartree mean-field approximation, the interacting QML is proportional to the noninteracting
QML, with a first-order correction in $U$. The interaction correction is therefore governed by the ratio
$Um/V$, which compares the Hartree reduction of the orbital
splitting with its bare value.
For our DMRG parameters, the small-$V/J$ condition is not satisfied, since $V=0.3-0.6, J=1$. Therefore
we apply Eq.~\eqref{eq:IQML-linear} as a theoretical estimate of the interacting QML, with $\bar n_C-\bar n_A$ given by DMRG for each $U$.

\subsection{Geometric contribution to the Majorana instability}

The noninteracting Lieb--Kitaev model supports a Majorana
wavefunction with a conventional superconducting component and
an additional geometric component
\cite{guoMajoranaZeroModes2025}.
For an open chain of length $\ell=(L-1)a$, we write their
long-distance form in terms of two decay lengths: $\xi_{\mathrm g} = 4 \xi_{\mathrm QM}$
and the Majorana decay length $\xi_{\mathrm s}$:
\begin{equation}
\begin{aligned}
\Psi_L(x)
&\simeq
\mathbf A_{L,\mathrm g}(x)e^{-x/\xi_{\mathrm g}}
+\mathbf A_{L,\mathrm s}(x)e^{-x/\xi_{\mathrm s}},\\
\Psi_R(x)
&\simeq
\mathbf A_{R,\mathrm g}(x)e^{-(\ell-x)/\xi_{\mathrm g}}
+\mathbf A_{R,\mathrm s}(x)e^{-(\ell-x)/\xi_{\mathrm s}}.
\end{aligned}
\label{eq:IQML-Majorana-wavefunctions}
\end{equation}

Here we choose the absolute value of the wave function. At zero chemical potential and for $0<|\Delta|<t$, the conventional
Kitaev decay length is
\begin{equation}
\xi_{\mathrm s}
=
\frac{2a}{
\ln[(t+|\Delta|)/(t-|\Delta|)]
}.
\end{equation}

Within this regime, the leading interaction correction can be
incorporated into the geometric part of the wavefunction as
\begin{equation}
\begin{aligned}
\Psi_{L,\mathrm g}(x)
&\simeq
\mathbf A_{L,\mathrm g}(x;U,V)
\exp\left[-\frac{x}{4\xi_{\mathrm{QML}}^{\mathrm{MF}}(U,V)}\right],\\
\Psi_{R,\mathrm g}(x)
&\simeq
\mathbf A_{R,\mathrm g}(x;U,V)
\exp\left[-\frac{\ell-x}
{4\xi_{\mathrm{QML}}^{\mathrm{MF}}(U,V)}\right].
\end{aligned}
\end{equation}

A longer geometric tail reduces the spatial separation between
the two boundary modes.
To illustrate this independently of their internal spinor
structure, consider normalized scalar envelopes
\begin{equation}
f_L(x)=\mathcal N e^{-x/\xi_{\mathrm g}},
\qquad
f_R(x)=\mathcal N e^{-(\ell-x)/\xi_{\mathrm g}},
\label{majorana}
\end{equation}
where
\begin{equation}
\mathcal N^2
=
\frac{2}{
\xi_{\mathrm g}(1-e^{-2\ell/\xi_{\mathrm g}})
}.
\end{equation}

As discussed in Ref.~\cite{Cheng2009MajoranaSplitting}, we let $\mathcal H_\infty$ be the effective BdG Hamiltonian on a
semi-infinite chain, where $\mathcal H_\infty|L\rangle=0$. The left Majorana mode $|L\rangle$ is a
zero-energy state under $\mathcal H_\infty$. Once an additional $\mathcal H_l$
is introduced as a boundary at $l$, $\Psi_L$ is no longer a zero-energy eigenstate and
$\mathcal H_l\Psi_L$ measures its deviation from zero energy. Therefore,
$t_{RL}=\langle R|\mathcal H_\ell|L\rangle$ is the
hybridization matrix element between the left and right Majorana end states:
\begin{equation}
\begin{aligned}
t_{RL}
&=\langle R|H|L\rangle
=\langle R|H_\infty|L\rangle
+\langle R|H_l|L\rangle
=\langle R|H_l|L\rangle,\\
\mathcal H_{\mathrm{edge}} & =\frac{i\varepsilon_M}{2}\gamma_L\gamma_R
=
\begin{pmatrix}
\langle L|H|L\rangle & \langle L|H|R\rangle\\
\langle R|H|L\rangle & \langle R|H|R\rangle
\end{pmatrix}
=
\begin{pmatrix}
0 & t_{RL}^{*}\\
t_{RL} & 0
\end{pmatrix}
\end{aligned}
\label{eq:Majorana-edge-coupling}
\end{equation}

Therefore, we can derive the hybridization  as
\begin{equation}
(H\psi)_j
=h\psi(j)+T\psi(j+1)+T^\dagger\psi(j-1), \qquad t_{RL}=-\psi_R^\dagger(M)\,T\,\psi_L(M+1).
\end{equation}

where $h$ and $T$ are the intracell and intercell BdG matrices.
In our Lieb--Kitaev model, these matrices are expressed in the Nambu basis $\bigl(
c_{jA},c_{jC},c_{jB},
c_{jA}^{\dagger},c_{jC}^{\dagger},c_{jB}^{\dagger}
\bigr)^{\mathsf T}$
\begin{equation}
h=
\begin{pmatrix}
V&0&-J&0&0&0\\
0&-V&-J&0&0&0\\
-J&-J&0&0&0&0\\
0&0&0&-V&0&J\\
0&0&0&0&V&J\\
0&0&0&J&J&0
\end{pmatrix},
\qquad
T=
\begin{pmatrix}
-t&0&0&-\Delta&0&0\\
0&-t&0&0&-\Delta&0\\
-J&-J&-t&0&0&-\Delta\\
\Delta&0&0&t&0&0\\
0&\Delta&0&0&t&0\\
0&0&\Delta&J&J&t
\end{pmatrix}.
\end{equation}

We denote the left and right Majorana wavefunctions by
\begin{equation}
\psi_\eta(j)
=
\begin{pmatrix}
u_{\eta A}(j)\\
u_{\eta C}(j)\\
u_{\eta B}(j)\\
v_{\eta A}(j)\\
v_{\eta C}(j)\\
v_{\eta B}(j)
\end{pmatrix},
\qquad 
-T\psi_L(M+1)
=
\begin{pmatrix}
t u_{LA}+\Delta v_{LA}\\
t u_{LC}+\Delta v_{LC}\\
J(u_{LA}+u_{LC})+t u_{LB}+\Delta v_{LB}\\
-\Delta u_{LA}-t v_{LA}\\
-\Delta u_{LC}-t v_{LC}\\
-\Delta u_{LB}-J(v_{LA}+v_{LC})-t v_{LB}
\end{pmatrix}_{j=M+1}.
\eta=L,R.
\label{eq:edge-wavefunction-components}
\end{equation}
Here $u_{\eta\alpha}(j)$ and $v_{\eta\alpha}(j)$ are the
particle and hole components of mode $\eta$ on orbital
$\alpha$ in cell $j$, respectively. $H$ belongs to the semi-infinite chain and the finite 
corrections contain $M$ unit cells. We now assume that the geometric component dominates
the left-mode tail at the opposite boundary, given by~\eqref{majorana}, and
the quantities
$\widetilde u_{L\alpha}$ and $\widetilde v_{L\alpha}$
are the remaining particle and hole amplitudes after
removing this exponential factor:
\begin{equation}
\begin{aligned}
u_{L\alpha}(M+1)
&\simeq
\widetilde u_{L\alpha}(M+1)e^{-Ma/\xi_{\mathrm g}},\\
v_{L\alpha}(M+1)
&\simeq
\widetilde v_{L\alpha}(M+1)e^{-Ma/\xi_{\mathrm g}}.
\end{aligned}
\end{equation}

Substituting these expressions into
Eq.~\eqref{eq:edge-wavefunction-components} gives
\begin{equation}
\begin{aligned}
t_{RL}\simeq e^{-Ma/\xi_{\mathrm g}}
\Biggl\{&
t\sum_{\alpha=A,C,B}
\Bigl[
u_{R\alpha}^{*}(M)\widetilde u_{L\alpha}(M+1)
-v_{R\alpha}^{*}(M)\widetilde v_{L\alpha}(M+1)
\Bigr]\\
&+\Delta\sum_{\alpha=A,C,B}
\Bigl[
u_{R\alpha}^{*}(M)\widetilde v_{L\alpha}(M+1)
-v_{R\alpha}^{*}(M)\widetilde u_{L\alpha}(M+1)
\Bigr]\\
&+J u_{RB}^{*}(M)
\Bigl[
\widetilde u_{LA}(M+1)+\widetilde u_{LC}(M+1)
\Bigr]\\
&-J v_{RB}^{*}(M)
\Bigl[
\widetilde v_{LA}(M+1)+\widetilde v_{LC}(M+1)
\Bigr]
\Biggr\}.
\label{eq:tRL-geometric-explicit}
\end{aligned}
\end{equation}

Here $u_{R}(M)$ is the right Majorana state at the right edge, so
all terms with $(t,\Delta ,J)$ remove this exponential factor below.
In the regime where
$\xi_{\mathrm g}\simeq4\xi_{\mathrm{QML}}$, this becomes
\begin{equation}
|t_{RL}|_{\mathrm{env}}
\propto
\exp\!\left[-\frac{L}{4\xi_{\mathrm{QML}}}\right].
\label{eq:tRL-QML-envelope}
\end{equation}

Here $L$ denotes the physical chain length. With the Hartree correction
to the interacting QML given by Eq.~\eqref{eq:IQML-linear}, we 
have
\begin{equation}
\begin{aligned}
t_{RL}\simeq 
\exp\Biggl\{
-\frac{L}{4\xi_{\mathrm{QML}}^{(0)}(V)}
\Biggl[1
&-U(\bar n_C-\bar n_A)
\left(
\frac1V+\frac{3V}{V^2+8J^2}
-\frac{2V}{V^2+4J^2}
\right)
\Biggr]
\Biggr\}.
\end{aligned}
\label{eq:tRL-Hartree-linear}
\end{equation}

In the small-$V/J$ regime, this simplifies to
\begin{equation}
t_{RL}
\simeq 
\exp\!\left[
-\frac{L}{\sqrt2 aJ}
\left\{V-U\bigl[\bar n_C(U,V)-\bar n_A(U,V)\bigr]\right\}
\right].
\label{eq:tRL-small-V-Hartree}
\end{equation}

The above result identifies a geometric mechanism that
weakens the finite-size protection of Majorana edge states.
A larger QML implies a longer penetration length of each
end mode and consequently a larger overlap between the
two ends. Since
$t_{RL}$ also represents the energy splitting of the two Majorana modes $\delta E_{P,\mathrm{env}}$,
increasing the QML weakens the exponential
suppression of the finite-size energy splitting.
The Hartree correction strengthens this mechanism when
$\bar n_C>\bar n_A$. In the small-$V/J$ regime,
\begin{equation}
\frac{\partial\ln\delta E_{P,\mathrm{env}}}{\partial U}
\simeq{}
\frac{L}{\sqrt2 aJ}
\left[
\bar n_C-\bar n_A
+U\frac{\partial(\bar n_C-\bar n_A)}{\partial U}
\right].
\end{equation}

An increasing positive quantity
$U(\bar n_C-\bar n_A)$ enhances the geometric contribution
to Majorana hybridization, although changes in the
prefactor and interference between decay components
can modify the splitting at individual parameters.
Also at fixed $J$, the
noninteracting QML increases as $V$ decreases.
For comparable boundary prefactors and orbital
polarizations, a smaller $V$ therefore produces a
larger initial overlap and a larger relative Hartree
correction, controlled by
$U(\bar n_C-\bar n_A)/V$.
Finite chains with longer QML will lose a degree of Majorana energy
degeneracy at weaker interactions and are therefore expected to exhibit a lower
TSC-ICDW phase transition point.

\section{Extraction of effective Luttinger parameters}
\label{app:luttinger_fits}

We organize the low-energy theory schematically as
$H=H_{\mathrm{LL,eff}}+H_{\Delta}+H_U$, comprising
a Gaussian LL reference, pairing and commensurate
locking.
In the usual density representation,
commensurate locking favors a lattice-compatible
density pattern, whereas a change in filling shifts
the preferred density wave vector. Their competition
can produce an incommensurate regime once the density
is no longer locked to the commensurate value,
expressed here as $Q=2k_Fa$.~\cite{mahyaehStudyPhaseDiagram2020,rahmaniPhaseDiagramInteracting2015,stoudenmireInteractionEffectsTopological2011a}

In the ICDW phase, the pairing perturbation is
irrelevant to the gapless mode; finite incommensurability
also suppresses Umklapp locking through its spatially
oscillating phase. At sufficiently long distances, the
remaining low-energy degrees of freedom are therefore
described by a Gaussian Tomonaga--Luttinger theory. We therefore follow
this theory, replacing $K$ with $K_{\mathrm{eff}}$ and
using an effective velocity $v_{\mathrm{eff}}$ from Eq.~\eqref{eq:app-K-v}. Below we show how $K$  
can be derived numerically from density correlation functions, 
density oscillations and bond oscillations under open boundary conditions.

\subsection{Luttinger-liquid preliminaries}
\label{app:ll_reference}

We use the field normalization
\begin{equation}
H_{\mathrm{LL}}
=
\frac{v}{2}\int dx\,
\left[
K(\partial_x\theta)^2
+\frac{1}{K}(\partial_x\varphi)^2
\right],
\label{eq:app_ll_hamiltonian}
\end{equation}
where $\partial_x\theta$ is conjugate to $\varphi$.
Although this Hamiltonian describes free bosonic
fields, $K$ and $v$ incorporate interaction effects;
the reference theory does not require
noninteracting microscopic fermions
\cite{giamarchi2004}.

To obtain the density operator, we decompose the
fermion field into slowly varying right- and
left-moving components near the two Fermi points
and bosonize them. The same-branch contributions
give the smooth density, while interference
between the two branches produces the leading
$2k_F$ harmonic:
\begin{equation}
\begin{split}
\rho(x)
={}&\rho_0
-\frac{1}{\sqrt{\pi}}\partial_x\varphi(x)\\
&+A\cos\left[
2k_Fx-2\sqrt{\pi}\varphi(x)
\right]+\cdots.
\end{split}
\label{eq:app_density_expansion}
\end{equation}
Here $A$ is a nonuniversal amplitude, and the
omitted terms include higher harmonics and
derivative corrections.

After integrating out the conjugate field, the
Euclidean action is
\begin{equation}
S_0
=
\frac{1}{2K}
\int dx\,d\tau
\left[
\frac{1}{v}(\partial_\tau\varphi)^2
+v(\partial_x\varphi)^2
\right].
\label{eq:app_ll_action}
\end{equation}

Starting from the Gaussian action, we evaluate
the functional integral to obtain the momentum-space
Green's function. An inverse Fourier transform then gives
the real-space Green function. Ultraviolet and
infrared cutoffs are introduced to regularize
the short- and long-wavelength divergences.

\begin{equation}
G_0(z)
=
\langle\varphi(z)\varphi(0)\rangle_0
=
-\frac{K}{2\pi}
\ln\frac{|z|}{a_0}
+\mathrm{const}.
\label{eq:app_phi_green}
\end{equation}

The leading density vertex is
$\mathcal V(x)=e^{i2\sqrt{\pi}\varphi(x)}$.
For a zero-mean Gaussian variable $X$, the
Gaussian identity
$\langle e^{i\lambda X}\rangle
=\exp[-\lambda^2\langle X^2\rangle/2]$
follows from the vanishing of all cumulants
beyond second order. Taking
$X=\varphi(x)-\varphi(0)$, its variance is
$\langle X^2\rangle_0=2[G_0(0)-G_0(x)]$.
Using the regularized logarithmic Green function,
this becomes
$\langle X^2\rangle_0
\simeq(K/\pi)\ln(|x|/a_0)$,
up to a cutoff-dependent constant. Therefore,
\begin{equation}
\begin{split}
\langle\mathcal V(x)\mathcal V^\dagger(0)\rangle_0
&=
\exp\left[
-2\pi
\langle(\varphi(x)-\varphi(0))^2\rangle_0
\right]\\
&=
\left(\frac{a_0}{|x|}\right)^{2K}.
\end{split}
\label{eq:app_vertex_bulk}
\end{equation}

\subsection{Density correlations}
\label{app:density_correlation}

The oscillatory term in
Eq.~\eqref{eq:app_density_expansion} carries
the physical wave vector $2k_F$.
For lattice coordinates $x=ai$, we introduce
the dimensionless wave vector $Q=2k_Fa$.
Within the spinless LL reference theory,
$k_F=\pi\rho_0$, so doping changes the density
harmonic directly through the mean density.
Writing the filling of the effective mode as
$\nu=a\rho_0=1/2+\delta\nu$, we obtain
$Q=2\pi\nu=\pi+2\pi\delta\nu
\quad (\mathrm{mod}\;2\pi).$ To connect this continuum description to the
orbital-resolved observable, we write
\begin{equation}
\begin{split}
n_{iB}\simeq{}&
n_{\mathrm{bg}}(i)
-\frac{Z_Ba}{\sqrt{\pi}}
\partial_x\varphi(x)\\
&+
A_B\cos\left[
Qi-2\sqrt{\pi}\varphi(x)+\eta_B
\right]+\cdots,
\end{split}
\label{eq:app_density_operator}
\end{equation}
where $Z_B$ describes the projection of the
smooth density mode onto the $B$ orbital,
while $A_B$ and $\eta_B$ are the amplitude
and phase of its oscillatory component.
Orbital
projection changes its coefficient without
changing its Gaussian scaling dimension,
provided that this vertex remains dominant.

We consider the connected density correlation
\begin{equation}
C_B(i,r)
=
\langle n_{iB}n_{i+r,B}\rangle
-\langle n_{iB}\rangle
 \langle n_{i+r,B}\rangle.
\label{eq:app_connected_definition}
\end{equation}
Substituting the density expansion into the
connected correlation function leaves two
leading contributions in the Gaussian bulk
limit. The first is the smooth-density term,
$(Z_B^2a^2/\pi)
\langle\partial_x\varphi(x)
\partial_{x'}\varphi(x')\rangle_0$.
Differentiating the Green function in
Eq.~\eqref{eq:app_phi_green} and setting
$x'-x=ar$ gives $-Z_B^2K/(2\pi^2r^2)$.
The second comes from the neutral part of the
product of the two cosine terms. It is proportional
to $\cos(Qr)$ multiplied by the vertex correlation
in Eq.~\eqref{eq:app_vertex_bulk}, and therefore
decays as $\cos(Qr)/r^{2K}$.
The mixed derivative--vertex terms and the
nonneutral vertex products vanish by Gaussian
charge neutrality. Absorbing the nonuniversal
prefactors into $A_C$, we obtain
\begin{equation}
C_B^{\mathrm{bulk}}(r)
\simeq
-\frac{D_B}{r^2}
+\frac{A_C\cos(Qr)}{r^{2K}},
\qquad
D_B=\frac{Z_B^2K}{2\pi^2}.
\label{eq:app_density_corr_form}
\end{equation}
The smooth component therefore decays as
$r^{-2}$, while the oscillatory envelope carries
the exponent $2K$. In the finite-window analysis,
the fitted parameter is interpreted as
$K_{\mathrm{eff}}$.

\subsection{Density oscillations}
\label{app:density_profile}

Taking the expectation value of the density
expansion ~\eqref{eq:app_density_expansion} and expressing the cosine through
complex exponentials gives
\begin{equation}
\langle\rho(x)\rangle
=
\rho_{\mathrm{bg}}(x)
+
A\,\operatorname{Re}
\left[
e^{i2k_Fx}
\left\langle
e^{-i2\sqrt{\pi}\varphi(x)}
\right\rangle
\right],
\label{eq:app_density_expectation}
\end{equation}
where the expectation value of the smooth
derivative term is included in
$\rho_{\mathrm{bg}}$.
The density envelope is therefore determined
by a one-point vertex expectation value.

For the open-boundary LL reference theory,
the density field is pinned at the two ends.
After incorporating the filling-dependent
background into $k_F$, we write
$\varphi(x,\tau)=\varphi_0+\delta\varphi(x,\tau)$,
where $\varphi_0$ is a constant fixed by the
boundary condition and contributes to the phase
of the density oscillation. The fluctuating field
then satisfies 
$\delta\varphi(0,\tau)=
\delta\varphi(\mathcal L,\tau)=0$.
It can consequently be expanded in standing
sine modes~\cite{cazalillaBosonizingOnedimensionalCold2004}: 
\begin{equation}
\delta\varphi(x,\tau)
=
\sqrt{\frac{2}{\mathcal L}}
\sum_{m=1}^{\infty}
\varphi_m(\tau)\sin(k_mx),
\qquad
k_m=\frac{m\pi}{\mathcal L},
\label{eq:app_obc_sine_modes}
\end{equation}

Substituting this expansion into the Gaussian
action~\eqref{eq:app_ll_action}, the orthogonality
of the sine modes eliminates cross terms between
different mode indices. Fourier transformation
in imaginary time then reduces the action to
independent quadratic modes. Evaluating their
Gaussian integrals gives the propagator
$Kv/(\omega^2+v^2k_m^2)$.
At zero temperature, integrating over frequency
and summing the modes with the short-distance
regulator $e^{-k_ma_0}$ yields the equal-position
variance

\begin{equation}
\langle\delta\varphi(x)^2\rangle_0
\simeq
\frac{K}{2\pi}
\ln\left[
\frac{2\mathcal L}{\pi a_0}
\sin\left(\frac{\pi x}{\mathcal L}\right)
\right].
\label{eq:app_obc_variance}
\end{equation}
Applying the Gaussian exponential identity
derived above, the magnitude of the vertex
expectation value becomes
\begin{equation}
\begin{split}
\left|
\left\langle
e^{-i2\sqrt{\pi}\varphi(x)}
\right\rangle_0
\right|
&=
e^{-2\pi\langle\delta\varphi(x)^2\rangle_0}\\
&\simeq
\left[
\frac{\pi a_0}
{2\mathcal L\sin(\pi x/\mathcal L)}
\right]^K.
\end{split}
\label{eq:app_obc_vertex}
\end{equation}
Thus, the sine-shaped finite-size envelope
follows from the open-boundary mode expansion,
rather than being introduced as an independent
fitting assumption
\cite{cazalillaBosonizingOnedimensionalCold2004}.

For $L$ unit cells, we place the effective
boundaries at $i=0$ and $i=L+1$, so that
$\mathcal L=(L+1)a$ and $x=ai$.
Defining
\begin{equation}
d(i)
=
\frac{L+1}{\pi}
\sin\left(\frac{\pi i}{L+1}\right),
\label{eq:app_chord_distance}
\end{equation}
and absorbing cutoff factors and orbital
projection amplitudes into $A_n$, we obtain
\begin{equation}
\langle n_{iB}\rangle
=
n_{\mathrm{bg}}(i)
+
\frac{A_n\cos(Qi+\phi_n)}
{[d(i)]^{K_{\mathrm{eff}}}}.
\label{eq:app_density_fit}
\end{equation}

\subsection{Bond oscillations}
\label{app:bond_profile}

We consider the normal bond observable
$b_i=\langle\hat b_i\rangle$, where
$\hat b_i=(c_{iB}^{\dagger}c_{i+1,B}
+\mathrm{h.c.})/2$.
Its low-energy expansion is naturally organized
about the bond center $X=a(i+1/2)$, with the
two lattice sites located at $x_\pm=X\pm a/2$.
Using the right- and left-moving decomposition
of the fermion field, we obtain
\begin{equation}
\begin{split}
c_{iB}^{\dagger}c_{i+1,B}\simeq a\big[
&|u_R|^2e^{ik_Fa}
R^\dagger(x_-)R(x_+)\\
+{}&|u_L|^2e^{-ik_Fa}
L^\dagger(x_-)L(x_+)\\
+{}&u_R^*u_Le^{-i2k_FX}
R^\dagger(x_-)L(x_+)\\
+{}&u_L^*u_Re^{i2k_FX}
L^\dagger(x_-)R(x_+)
\big],
\end{split}
\label{eq:app_bond_expansion}
\end{equation}
where $u_R$ and $u_L$ are the orbital projection
coefficients at the two Fermi points.
The first two terms describe the smooth
same-branch contribution, whereas the last
two terms generate the oscillatory component.

We expand the slowly varying fields about $X$,
using
$\psi(x_\pm)=\psi(X)\pm(a/2)\partial_x\psi(X)
+(a^2/8)\partial_x^2\psi(X)+\cdots$.
At leading order, the cross-branch contribution
is proportional to
$e^{-i2k_FX}R^\dagger L+\mathrm{h.c.}$.
Bosonizing $R^\dagger L$, the dual-field
contributions cancel and leave
$R^\dagger L\propto e^{i2\sqrt{\pi}\varphi}$.
The oscillatory and smooth parts of the bond operator consequently
take the forms

\begin{equation}
\begin{aligned}
\hat b_i^{\mathrm{osc}}
\simeq{}&
\mathcal A_b
\cos\left[
2k_FX-2\sqrt{\pi}\varphi(X)+\eta_b
\right],\\
\hat b^{\mathrm{sm}}(X)
\simeq{}&
b_0+\beta_1\partial_x\varphi(X)\\
&+\beta_\varphi:
(\partial_x\varphi(X))^2:
+\beta_\theta:
(\partial_x\theta(X))^2:
\end{aligned}
\label{eq:app_bond_components}
\end{equation}

The cutoff dependence, orbital coefficients,
and Klein-factor phase are absorbed into
$\mathcal A_b$ and $\eta_b$. Normal ordering subtracts the
short-distance bulk contribution.

To evaluate this correction, we write
$\varphi=\varphi_0+\delta\varphi$ and
$\theta=\theta_0+\delta\theta$, with the
filling-dependent background already separated.
For the open-boundary LL reference, the density
field has fixed endpoint values, so that
$\delta\varphi(0)=\delta\varphi(\mathcal L)=0$.
The corresponding no-current condition imposes
$\partial_x\theta=0$ at both boundaries.
We therefore expand $\delta\varphi$ in normalized
sine modes and $\delta\theta$ in normalized
cosine modes, with $k_m=m\pi/\mathcal L$.

\begin{equation}
\begin{aligned}
\delta\varphi(x,\tau)
&=
\sqrt{\frac{2}{\mathcal L}}
\sum_{m=1}^{\infty}
\varphi_m(\tau)\sin(k_mx),\\
\delta\theta(x,\tau)
&=
\sqrt{\frac{2}{\mathcal L}}
\sum_{m=1}^{\infty}
\theta_m(\tau)\cos(k_mx),
\end{aligned}
\label{eq:app_obc_dual_modes}
\end{equation}

Substituting the mode expansions into the
Gaussian theory, spatial integration eliminates
cross terms between different mode indices by
orthogonality. The resulting normal modes are
independent Gaussian oscillators.
Fourier transformation in imaginary time gives
the mode propagators
$Kv/(\omega^2+v^2k_m^2)$ and
$v/[K(\omega^2+v^2k_m^2)]$ for
$\varphi_m$ and $\theta_m$, respectively.
At zero temperature, integrating over frequency
gives the equal-time variances
$\langle\varphi_m^2\rangle_0=K/(2k_m)$ and
$\langle\theta_m^2\rangle_0=1/(2Kk_m)$.

Differentiating the mode expansions introduces
a factor $k_m$ and changes the sine basis of
$\delta\varphi$ into a cosine basis, while
changing the cosine basis of $\delta\theta$
into a sine basis. The constant backgrounds
do not contribute to these derivatives.
Combining the squared mode derivatives with
the equal-time variances, the normalization
factor $2/\mathcal L$, and the ultraviolet
regulator $e^{-k_ma_0}$ yields
\begin{equation}
\begin{aligned}
\langle(\partial_x\varphi(X))^2\rangle_0
&=
\frac{K\pi}{\mathcal L^2}
\sum_{m=1}^{\infty}
m e^{-m\epsilon}
\cos^2\left(\frac{m\pi X}{\mathcal L}\right),\\
\langle(\partial_x\theta(X))^2\rangle_0
&=
\frac{\pi}{K\mathcal L^2}
\sum_{m=1}^{\infty}
m e^{-m\epsilon}
\sin^2\left(\frac{m\pi X}{\mathcal L}\right),
\end{aligned}
\label{eq:app_bond_gradient_sums}
\end{equation}
where $\epsilon=\pi a_0/\mathcal L$.

Evaluating the mode sums with
$\sum_{m\ge1}m z^m=z/(1-z)^2$ and absorbing
position-independent contributions into $b_0$
leaves an inverse-square boundary correction
from the quadratic-gradient terms.
Combining these contributions yields
\begin{equation}
b_i
\simeq
b_0+\frac{C_2}{[d_b(i)]^2}
+
\frac{
A_b\cos[Q(i+1/2)+\phi_b]
}{
[d_b(i)]^{K_{\mathrm{eff}}}
},
\label{eq:app_bond_full_form}
\end{equation}
where $d_b(i)=d(i+1/2)$~\eqref{eq:app_chord_distance}, $Q=2k_Fa$, and
$C_2=\sin(Q/2)(K_{\mathrm{eff}}-K^{-1}_{\mathrm{eff}})/(16\pi)$. 
The exponent is interpreted as
$K_{\mathrm{eff}}$ in a finite fitting window.

\section{Numerical Strategy}
\label{numeriacal details}

The interacting ground state is computed by finite-system DMRG with open boundary conditions using ITensor
\cite{fishman2022}. The three-orbital chain is mapped to a one-dimensional matrix-product-state sequence
$(A_1,B_1,C_1,A_2,B_2,C_2,\dots,A_L,B_L,C_L)$. We choose the parameters $t=0.03$, $V=0.6$, $J=1$, and
$\Delta=0.018$ and restrict the system to half filling for this spinless model. The source calculations
emphasize careful sweep scheduling and convergence control, which is essential because the entanglement
rises substantially in parts of the repulsive regime and because flat-band systems generally exhibit
a large degree of degeneracy. Each parameter point is initialized independently from the empty-site
product state and optimized through a prescribed
sequence of 80 full sweeps. The maximum bond dimension
is increased progressively through
$20$, $40$, $80$, $160$, $320$, $640$, and $1280$,
while the truncation cutoff is tightened from $10^{-4}$
through $10^{-6}$ and $10^{-8}$. A density-matrix noise term is included during the initial optimization, with its coefficient gradually
reduced from $10^{-5}$ to 0. A custom observer records the total energy and the
von Neumann entanglement entropy after every full sweep. Convergence is assessed by requiring both the absolute
change in the total energy and the maximum change in the
entanglement entropy over all intercell cuts between the
final two sweeps to be smaller than $10^{-7}$. 
Table~\ref{tab:DMRG-observables} summarizes the observables
evaluated from the DMRG ground-state MPS and the quantities
subsequently extracted from them.

\begin{table*}[t]
\centering
\caption{
Observables and analysis procedures used in the DMRG calculations.
Here $i,j$ label unit cells, $\alpha,\beta\in\{A,B,C\}$
label orbitals, and $L=N/3$.
Expectation values are evaluated in the normalized
ground-state MPS $|\Psi_0\rangle$.
}
\label{tab:DMRG-observables}
\footnotesize
\renewcommand{\arraystretch}{1}
\begin{tabular}{@{}ll@{}}
\hline \qquad \qquad \qquad \qquad
\textbf{Observable and extraction} & \qquad \qquad \qquad \qquad \qquad \qquad \qquad \qquad
\textbf{Definition} \\[3pt]
\hline
\parbox[t]{0.43\textwidth}{\textbf{Density profile}\par\smallskip
\strut}
&
\parbox[t]{0.53\textwidth}{$n_\alpha(i)=\langle n_{i\alpha}\rangle$\strut}\\[4pt]

\parbox[t]{0.43\textwidth}{\textbf{Ground-state filling}\par\smallskip
\strut}
&
\parbox[t]{0.53\textwidth}{$\bar n=N^{-1}\sum_{i,\alpha}n_\alpha(i)$\strut}\\[4pt]

\parbox[t]{0.43\textwidth}{\textbf{Normal and anomalous single-particle correlation functions}\par\smallskip
\strut}
&
\parbox[t]{0.53\textwidth}{$G_{\alpha\beta}(i,j)
=\langle c^\dagger_{i\alpha}c_{j\beta}\rangle$,
\par\smallskip
$F_{\alpha\beta}(i,j)
=\langle c_{i\alpha}c_{j\beta}\rangle$\strut}\\[4pt]

\parbox[t]{0.43\textwidth}{\textbf{Bond profile}\par\smallskip
\strut}
&
\parbox[t]{0.53\textwidth}{$B_\alpha(i)
=\langle c^\dagger_{i\alpha}c_{i+1,\alpha}\rangle$\strut}\\[4pt]

\parbox[t]{0.43\textwidth}{\textbf{Pair profile}\par\smallskip
\strut}
&
\parbox[t]{0.53\textwidth}{$\langle P_\alpha(i)\rangle$,
\par\smallskip
$P_\alpha(i)=c_{i\alpha}c_{i+1,\alpha}$\strut}\\[4pt]

\parbox[t]{0.43\textwidth}{\textbf{Connected density correlation}\par\smallskip
\strut}
&
\parbox[t]{0.53\textwidth}{$C^n_{\alpha\beta}(i,j)
=\langle n_{i\alpha}n_{j\beta}\rangle
-n_\alpha(i)n_\beta(j)$\strut}\\[4pt]

\parbox[t]{0.43\textwidth}{\textbf{Connected pair correlation}\par\smallskip
\strut}
&
\parbox[t]{0.53\textwidth}{$C^p_{\alpha\beta}(i,j)
=\langle P_\alpha^\dagger(i)P_\beta(j)\rangle
-\langle P_\alpha^\dagger(i)\rangle
 \langle P_\beta(j)\rangle$\strut}\\[4pt]

\parbox[t]{0.43\textwidth}{\textbf{Connected bond correlation}\par\smallskip
\strut}
&
\parbox[t]{0.53\textwidth}{$C^{b}_{\alpha\beta}(i,j)
=\langle B^\dagger_\alpha(i)B_\beta(j)\rangle
-\langle B^\dagger_\alpha(i)\rangle
 \langle B_\beta(j)\rangle$\strut}\\[4pt]

\parbox[t]{0.43\textwidth}{\textbf{Bulk-averaged correlations}\par\smallskip
\strut}
&
\parbox[t]{0.53\textwidth}{$\overline C_\alpha(r)
=\frac{1}{N_r}\sum_{i\in\mathcal W_r}
C_{\alpha\alpha}(i,i+r)$\strut}\\[4pt]

\parbox[t]{0.43\textwidth}{\textbf{Entanglement entropy}\par\smallskip
\strut}
&
\parbox[t]{0.53\textwidth}{$S(\ell)=-\sum_m s_m^2(\ell)\ln s_m^2(\ell)$\strut}\\[4pt]

\parbox[t]{0.43\textwidth}{\textbf{Effective central charge}\par\smallskip
\strut}
&
\parbox[t]{0.53\textwidth}{$S(\ell)=
\frac{c_{\mathrm{eff}}}{6}
\ln\!\left[
\frac{2L}{\pi}\sin\!\left(\frac{\pi\ell}{L}\right)
\right]+s_0$\strut}\\[4pt]

\parbox[t]{0.43\textwidth}{\textbf{Fourier Q}\par\smallskip
\strut}
&
\parbox[t]{0.53\textwidth}{$F(r)=r^2\overline C_B(r),Y(r)=F(r+1)-F(r)$\strut}\\[4pt]

\parbox[t]{0.43\textwidth}{\textbf{Correlation ratio}\par\smallskip
\strut}
&
\parbox[t]{0.53\textwidth}{$R_F=1-
|F_Y(Q-2\pi/L)|/|F_Y(Q)|$\strut}\\[4pt]
\hline
\end{tabular}
\end{table*}

In particular, in processing correlation functions
for each separation $r$, the disconnected
contribution is subtracted separately for
each starting position in the bulk before averaging:
\begin{equation}
\overline C_B(r)
=
\frac{1}{21}\sum_{i=20}^{40}
\left[
\langle n_{iB}n_{i+r,B}\rangle
-\langle n_{iB}\rangle
 \langle n_{i+r,B}\rangle
\right].
\label{eq:app_numerical_connected_average}
\end{equation}

At each interaction strength, the wave vector
is fixed using the total ground-state occupation,
$\mathcal N(U)=\sum_{i,\alpha}\langle n_{i\alpha}\rangle$:
\begin{equation}
Q_N(U)
=
\arccos\left[
\cos\left(\frac{2\pi\mathcal N(U)}{L}\right)
\right].
\label{eq:app_numerical_Q}
\end{equation}

This folds the occupation-derived wave vector
into $[0,\pi]$ and gives $Q_N=\pi$ at
$\mathcal N=3L/2$.
The same $Q_N$ is used for all fitting windows
at a given $U$. For each $U$, the total
ground-state occupation is obtained by summing
the local densities, which determines the
corresponding $Q_N$.

Equation~\eqref{eq:app_density_corr_form} contains
a rapidly decaying smooth contribution and
an oscillatory component associated with
charge-ordering correlations.
To emphasize the oscillatory contribution and reduce calculation error,
we define $F(r)=r^2\overline C_B(r)$.
Within the leading asymptotic expression,
the rapidly decaying term becomes a constant and
therefore cancels in the first difference,
$Y(r)=F(r+1)-F(r)$.
We select the windows $14\le r\le20$ and
$15\le r\le20$ to reduce the influence of
short-distance lattice corrections and residual
background terms not removed by differencing. The numerical fitting function is
\begin{equation}
\begin{split}
Y_{\mathrm{fit}}(r)=A_C\big[
&(r+1)^{2-2K_{\mathrm{eff}}}
\cos(Q_N(r+1))\\
&-r^{2-2K_{\mathrm{eff}}}\cos(Q_Nr)
\big].
\end{split}
\label{eq:app_numerical_difference_fit}
\end{equation}

where the oscillatory phase $\phi_b$ in
$\cos(Qr+\phi_b)$ is fixed to zero. We retain the signs of the data throughout
and do not take their absolute values or
logarithms. This avoids singular behavior
at zero crossings of the oscillation.

We fit the orbital density
$n_i=\langle n_{iB}\rangle$ and the normal bond
profile 
$b_i=\operatorname{Re}
\langle c_{iB}^{\dagger}c_{i+1,B}\rangle$
using Eqs.~\eqref{eq:app_density_fit} and~\eqref{eq:app_bond_full_form}. 
At each $U$, both observables use the same
occupation-derived wave vector $Q_N(U)$
defined in Eq.~\eqref{eq:app_numerical_Q}.
The wave vector is fixed throughout the
optimization and shared by all fitting windows. To reduce boundary-scale corrections and
examine window dependence, we retain
$s<i\le L-s$ with $s=8,10,12,15$.
For $L=80$, these choices give the windows
$9$--$72$, $11$--$70$, $13$--$68$, and
$16$--$65$, respectively.
The same windows are applied at every $U$
and to both observables.

\begin{figure*}[!htbp]
\centering

\includegraphics[width=\textwidth]
{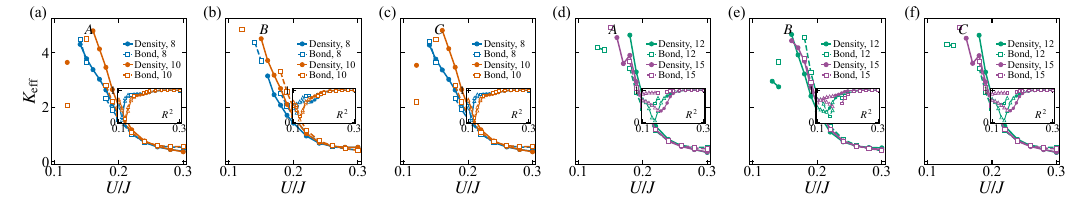}
\caption{
Orbital-resolved $K_{\mathrm{eff}}$ extracted from density and bond
profiles for $N=240$.
Panels (a)--(c) show orbitals $A$, $B$, and $C$ for $s=8,10$;
panels (d)--(f) show the same orbitals for $s=12,15$.
The fitting windows satisfy $s<i\leq L-s$, with $L=80$.
Insets show the corresponding coefficients of determination $R^2$.
}
\label{fig:S1}

\vspace{6pt}

\includegraphics[width=\textwidth]
{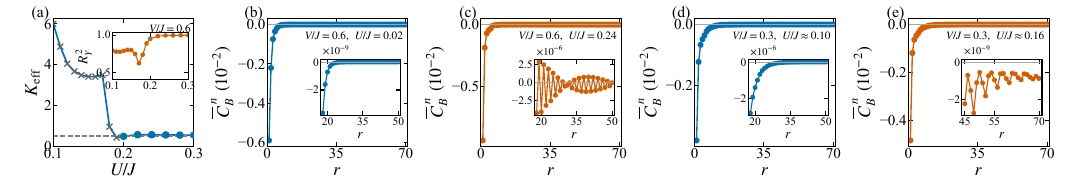}
\caption{
Density-correlation analysis.
(a) Analysis of $K_{\mathrm{eff}}$ at $V/J=0.6$, using
the $10\leq r\leq30$ window.
(b)--(e) Raw connected $B$-orbital density correlations at
$(V/J,U/J)=(0.6,0.02)$, $(0.6,0.24)$,
$(0.3,0.102105)$, and $(0.3,0.161053)$, respectively.
The latter two interaction strengths are labeled approximately
as $0.10$ and $0.16$ in the panels.
}
\label{fig:S2}

\vspace{6pt}

\includegraphics[width=\textwidth]
{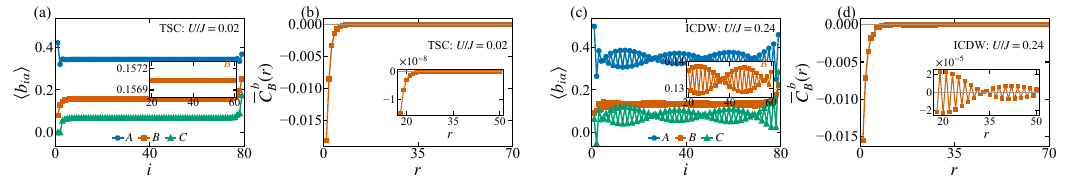}
\caption{
Bond observables.
(a),(b) TSC representative at $U/J=0.02$;
(c),(d) ICDW representative at $U/J=0.24$.
(a),(c) Orbital-resolved bond profiles $\langle b_{i\alpha}\rangle$,
where $b_{i\alpha}=c^\dagger_{i\alpha}c_{i+1,\alpha}
+\mathrm{H.c.}$.
(b),(d) Connected $B$-orbital bond correlations
$\overline{C}_B^{\,b}(r)
=\overline{\langle b_{iB}b_{i+r,B}\rangle
-\langle b_{iB}\rangle\langle b_{i+r,B}\rangle}$,
averaged over bond starts within $5\leq i,i+r\leq74$.
}
\label{fig:S3}

\vspace{6pt}

 \centering
 \includegraphics[width=0.49\textwidth]{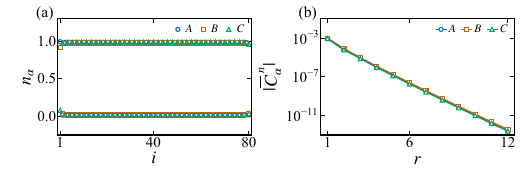}
 \caption{
 CDW phase.
 (a) Orbital occupations $n_\alpha(i)$.
 (b) Density correlations exhibit rapid exponential decay.
 }
 \label{fig:S4}

\end{figure*}

The fitting procedure is not restricted to the $B$ orbital
discussed in the main text. Applying the same analysis to
the $A$ and $C$ density and bond profiles also yields
orbital-resolved estimates of $K_{\mathrm{eff}}$, as shown
in Figs.~\ref{fig:S1} and~\ref{fig:S2}. When the incommensurate modulations on all three orbitals
originate from the same gapless density mode, their leading
density vertices share a common scaling dimension.
Orbital projection changes the nonuniversal coefficients
$Z_\alpha$, oscillation amplitudes, and phases, but not
the leading exponent.
The density and bond profiles are therefore expected to
yield a common $K_{\mathrm{eff}}$ in the asymptotic scaling
regime.

\FloatBarrier

\end{document}